\documentclass[journal]{IEEEtran}
\usepackage{amsmath,amsfonts}
\usepackage{algorithmicx}
\usepackage{algpseudocode}
\usepackage{algorithm}
\usepackage{array}
\usepackage{color}
\usepackage{textcomp}
\usepackage{stfloats}
\usepackage{url}
\usepackage{verbatim}
\usepackage{graphicx}
\usepackage{subcaption}
\usepackage{cite}
\usepackage{hyperref}
\hypersetup{
	colorlinks=false,
	linkcolor=cyan,
	filecolor=blue,      
	urlcolor=red,
	citecolor=green,
}
\usepackage{booktabs} % 用于专业的三线表格式
\usepackage{amsmath}  % 提供数学符号支持
\usepackage{amssymb}
\usepackage{float}
\makeatletter
\let\myorg@bibitem\bibitem  %把 LaTeX 原始的 \bibitem 命令保存为 \myorg@bibitem，防止后续重定义后丢失
\def\bibitem#1#2\par{%   重定义 \bibitem
	\@ifundefined{bibitem@#1}{%
		\myorg@bibitem{#1}#2\par
	}{%
		\begingroup
		\color{\csname bibitem@#1\endcsname}%
		\myorg@bibitem{#1}#2\par
		\endgroup
	}%
}
\makeatother
\begin{document}

\title{Knowledge Distillation Driven Semantic NOMA for Image Transmission with Diffusion Model}

\author{Qifei Wang, Zhen Gao,~\IEEEmembership{Senior Member,~IEEE}, Shuo Sun, Zhijin Qin,~\IEEEmembership{Senior Member,~IEEE}, Xiaodong Xu,~\IEEEmembership{Senior Member,~IEEE} and Meixia Tao,~\IEEEmembership{Fellow,~IEEE} 
        % <-this % stops a space
\thanks{The work of Z. Gao was supported in part by the Natural Science Foundation of China (NSFC) under Grant 62471036, in part by Shandong Province Natural Science Foundation under Grant ZR2025QA30, in part by Beijing Natural Science Foundation under Grants L242011, QY24167, QY25256, QY25257 (Undergraduate ”Qiyan” Research Program). Dr Zhijin Qin is supported by the National Natural Science Foundation of China under Grant No. 62293484, and Beijing Natural Science Foundation (F251001). The work by M. Tao was supported by the  National Natural Science Foundation of China (NSFC) under Grant 62125108 and by the National Science and Technology Major Project - Mobile Information Networks under Grant 2024ZD1300700.}
\thanks{Qifei Wang and Shuo Sun is with the School of Information and Electronics, Beijing Institute of Technology, Beijing 100081, China (e-mail: \href{mailto:qfwang@bit.edu.cn}{qfwang@bit.edu.cn}, \href{mailto:sunshuo2002@bit.edu.cn}{sunshuo2002@bit.edu.cn}).}%
\thanks{Zhen Gao (\textit{corresponding author}) is with  Beijing Institute of Technology (BIT), Zhuhai 519088, China, also with the State Key Laboratory of CNS/ATM, Beijing 100081, China, also with the MIIT Key Laboratory of Complex-Field Intelligent Sensing, Beijing 100081, China, also with the Advanced Technology Research Institute, BIT, Jinan 250307, China, and also with the Yangtze Delta Region Academy, BIT, Jiaxing 314019, China (e-mail: \href{mailto:gaozhen16@bit.edu.cn}{gaozhen16@bit.edu.cn}).} 
\thanks{Zhijin Qin is with the Department of Electronic Engineering, Tsinghua University, Beijing 100084, China. She is also with the State Key Laboratory of Space Network and Communications and the Beijing National Research Center for Information Science and Technology, Beijing 100084, China (e-mail: \href{mailto:qinzhijin@tsinghua.edu.cn}{qinzhijin@tsinghua.edu.cn}).}
\thanks{Xiaodong Xu is with the State Key Laboratory of Networking and Switching Technology, Beijing University of Posts and Telecommunications, Beijing 100876, China, and also with Peng Cheng Laboratory, Shenzhen 518066, China (e-mail: \href{mailto:xuxiaodong@bupt.edu.cn}{xuxiaodong@bupt.edu.cn}).}
\thanks{Meixia Tao is with the School of Information Science and Electronic Engineering at Shanghai Jiao Tong University, Shanghai 200240, China (e-mail: \href{mailto:mxtao@sjtu.edu.cn}{mxtao@sjtu.edu.cn}).}
\vspace*{-5mm}}



\maketitle

\begin{abstract}
As a promising 6G enabler beyond conventional bit-level transmission, semantic communication can considerably reduce required bandwidth resources, while its combination with multiple access requires further exploration. This paper proposes a knowledge distillation-driven and diffusion-enhanced (KDD) semantic non-orthogonal multiple access (NOMA), named KDD-SemNOMA, for  multi-user uplink wireless image transmission. Specifically, to ensure robust feature transmission across diverse transmission conditions, we firstly develop a ConvNeXt-based deep joint source and channel coding architecture with enhanced adaptive feature module. This module incorporates signal-to-noise ratio  and channel state information to dynamically adapt to additive white Gaussian noise and Rayleigh fading channels. Furthermore, to improve image restoration quality without inference overhead, we introduce a two-stage knowledge distillation strategy, i.e., a teacher model, trained on interference-free orthogonal transmission, guides a student model via feature affinity distillation and cross-head prediction distillation. Moreover, a diffusion model-based refinement stage leverages generative priors to transform initial SemNOMA outputs into high-fidelity images with enhanced perceptual quality. Extensive experiments on CIFAR-10 and FFHQ-256 datasets demonstrate superior performance over state-of-the-art methods, delivering satisfactory reconstruction performance even at extremely poor channel conditions. These results highlight the advantages in both pixel-level accuracy and perceptual metrics, effectively mitigating interference and enabling high-quality image recovery.

\end{abstract}

\begin{IEEEkeywords}
Semantic communication, non-orthogonal multiple access, channel adaptation, knowledge distillation, diffusion model.
\end{IEEEkeywords}
%\vspace{-3mm}
\section{Introduction}
\IEEEPARstart{W}{ith} the gradual advancement of 6G wireless networks, wireless communication is entering a transformative era of technological innovation. Future 6G networks will not only accommodate an explosive growth in data traffic but also meet escalating user demands for intelligent and efficient connection services {\color{black}\cite{wang2023road, gaocompressive, kecompressive}}. %In this context, traditional bit-oriented communication paradigm has gradually revealed limitations, which has promoted the exploration of next-generation technologies. 
In this context, traditional bit-oriented communication paradigm has gradually revealed limitations{\color{black}\cite{qingeneralized}}. Among the promising candidates for 6G, semantic communication has garnered significant attention due to its potential to transcend the constraints of conventional bit-level transmission \cite{gunduz2022beyond}. By integrating artificial intelligence, semantic communication shifts transmission objectives from bit-level to semantic-level information delivery, demonstrating superior transmission efficiency over traditional separated source-channel coding methods in bandwidth-constrained and complex channel environments {\color{black}\cite{bourtsoulatze2019deep}}. 

{\color{black}Semantic communication has achieved remarkable progress in single-user  scenarios. On the one hand, numerous studies have adopted end-to-end deep joint source-channel coding (DeepJSCC) frameworks to enable efficient semantic transmission of diverse modalities \cite{bourtsoulatze2019deep, xie2021deep}. On the other hand, another line of research has integrated generative models to enhance semantic reconstruction quality, particularly under challenging channel conditions \cite{duenhancing, liucross, yingenerative, yangofdm}.The paper \cite{yingenerative} proposes a generative video semantic communication framework, the work \cite{yangofdm} incorporates adversarial loss into a DeepJSCC-based orthogonal frequency division multiplexing (OFDM) system to enhance the perceptual quality of reconstructed images. These generative approaches have demonstrated superior performance in perceptual metrics, effectively compensating for the limitations of traditional distortion-based measures. Further improvements in dynamic adaptability and robustness are investigated in \cite{xu2021wireless}, which introduces attention modules for signal-to-noise ratio (SNR) adaptation in DeepJSCC codecs, and \cite{dai2022nonlinear}, which employs learnable entropy models for variable-length joint source-channel coding (JSCC).

In contrast, research on multi-user semantic communication remains in its infancy, with design complexity far exceeding that of single-user systems. In multi-user settings, semantic interference, limited transmission resources, and inter-user channel contention emerge as primary bottlenecks. The authors of \cite{xie2022task} propose task-oriented frameworks for multi-user semantic communication, while the authors of \cite{luo2022multimodal} explores the feasibility of multimodal semantic fusion. The paper \cite{wu2024deep} combines beamforming algorithm and semantic coding in the downlink massive multiple-input multiple-output (MIMO) scenario to achieve multi-user image downlink semantic transmission. Nonetheless, these approaches predominantly rely on orthogonal resource allocation, resulting in a sharp increase in physical-layer overhead as the number of users grows. To address this, non-orthogonal multiple access (NOMA) concepts have been introduced. For instance, the paper \cite{yilmaz2023distributed} proposes the end-to-end (E2E) uplink DeepJSCC-NOMA network for wireless image transmission. These works \cite{qiaotoken,li2023non, zhang2024interference, bo2025deep} also design NOMA based semantic communication system with some interference management optimization methods. However, these methods are still insufficient in terms of interference mitigation in NOMA scenarios.}

%However, designing end-to-end semantic communication systems for multi-user scenarios face significant challenges: adapting to diverse channel conditions (e.g., additive white Gaussian noise [AWGN] channel and Rayleigh fading channel) \cite{luo2022multimodal}, mitigating non-orthogonal feature interference {\color{red}\cite{liang2024orthogonal}}, and enhancing perceptual quality under strict bandwidth constraints \cite{luo2022semantic}. Specifically, non-orthogonal transmission, where multiple users’ semantic features share the same time-frequency resources, significantly exacerbates interference, while diverse channels and low compression ratios degrade signal reconstruction quality.  {\color{red}Conventional resource management approaches (e.g., priority-aware scheduling \cite{huanglow}) fail to effectively handle massive communication with severe interference.}
%{\color{red} This paper presents a novel solution that synergistically combines knowledge distillation and diffusion models to overcome these critical challenges. Our approach demonstrates remarkable capabilities in interference suppression and robust image quality preservation under adverse channel conditions.}  
\subsection{Motivation and Contributions}
Although prior studies have leveraged deep learning techniques to design multi-user semantic communication systems, significant challenges persist in the context of uplink semantic NOMA transmission scenarios. These include adapting to dynamic channel conditions such as additive white Gaussian noise (AWGN) and Rayleigh fading channels, inadequate management of non-orthogonal semantic feature interference, and limited enhancements in multi-user perceptual image quality under strict bandwidth constraints. Dynamic channels degrade transmission performance, interference from superimposed different users' signals reduces pixel-level fidelity, and high compression ratios and poor channel conditions impair perceptual metrics. These limitations hinder scalable 6G network deployment. Addressing these challenges in semantic NOMA communication, this paper proposes a novel framework that enhances image recovery performance across both pixel-level fidelity and perceptual quality. The main contributions of this work are summarized as follows:

\begin{itemize}
	\item \textbf{Non-orthogonal semantic transmission framework}: We propose a semantic NOMA framework (SemNOMA) for multi-user image uplink transmission, where a ConvNeXt-based DeepJSCC network is designed and an enhanced channel feature adaptation module is incorporated to ensure robust performance across AWGN and Rayleigh fading channels.
	\item \textbf{Knowledge distillation for non-orthogonal transmission optimization}: We innovatively incorporate knowledge distillation scheme into the SemNOMA framework, called KD-SemNOMA, where an orthogonal transmission-based teacher model is utilized to guide SemNOMA training via feature affinity and cross-head distillation. Simulations demonstrate that the introduction of the knowledge distillation framework yields a peak signal-to-noise ratio (PSNR) performance gain of approximately $0.2 \sim 0.8$dB compared with DeepJSCC-NOMA \cite{yilmaz2023distributed}, significantly enhancing the SemNOMA restoration performance without incurring additional computational overhead during inference.
	\item \textbf{Diffusion model-enhanced image refinement}: We further propose a two-stage reconstruction approach, called KDD-SemNOMA, where initial KD-SemNOMA (the first stage) outputs are refined by exploiting the error contraction property and generative priors of pretrained diffusion models  (the second stage), leading to substantial improvements in perceptual quality and detail fidelity. Extensive experiments on different datasets have shown that our method significantly outperforms previous multi-user DeepJSCC NOMA methods in terms of image pixel distortion and perceptual metrics, demonstrating its capability to preserve the original transmitted image semantics. 
\end{itemize}

The rest of this paper is organized as follows. In Section \ref{System Model}, we introduce the system model and describe the overall structure of SemNOMA framework. In Section \ref{KD}, we introduce the knowledge distillation strategy to optimize SemNOMA. In Section \ref{Diffusion}, we show that the diffusion prior can further recover the details of the reconstructed  images. Numerical simulation results are provided in Section \ref{Simulation}, followed by the conclusion of this work in Section \ref{Conclusion}. 

\textit{Notation}: $\mathbb{R}$ and $\mathbb{C}$ denote the real and complex number sets, respectively. $\mathcal{N}(\mu, \sigma^2)$ and $\mathcal{C N}\left(\mu, \sigma^2\right)$ denote the Gaussian function and the complex form of the Gaussian function with mean $\mu$ and variance $\sigma^2$, respectively. $\left\| \cdot \right\|_1$ denotes the $L_1$-norm (sum of absolute values). $[\cdot]$ denotes indexing into a multi-dimensional array or tensor. Superscript $(\cdot)^{\mathrm{T}}$ denotes the transpose. $\mathbf{I}_m$ denotes the $m \times m$ identity matrix. $|\mathbf{A}|_c$ denotes the number of elements in the matrix $\mathbf{A}$.

\section{Related Works}
{\color{black} This section reviews recent advances in semantic communication, focusing on three key areas: multi-user transmission techniques, knowledge distillation applications in semantic communication, and diffusion models for semantic enhancement.}
{\color{black}\subsection{DeepJSCC in Multi-User Semantic Communication}} \label{musemcom}
%Semantic communication has achieved remarkable progress in single-user  scenarios, with numerous studies employing deep joint source-channel coding (DeepJSCC) to enable efficient semantic transmission of diverse modalities \cite{bourtsoulatze2019deep, xie2021deep}. %including images \cite{bourtsoulatze2019deep}, text \cite{xie2021deep}, and audio \cite{zhou2023speech}. 
%The paper \cite{xu2021wireless} introduces attention modules to adaptive signal-to-noise ratio (SNR) in DeepJSCC codec, and the work \cite{dai2022nonlinear} introduces learnable entropy models to implement joint source-channel coding (JSCC) with variable-length transmission mechanism. These works underscore the potential of semantic communication in terms of dynamic performance and robustness. However, research on multi-user semantic communication remains in its infancy, with design complexity far exceeding that of single-user systems. 
In multi-user settings, semantic interference, limited transmission resources, and inter-user channel contention emerge as primary bottlenecks. The paper \cite{yilmaz2023distributed} proposes the end-to-end (E2E) uplink DeepJSCC-NOMA network for wireless image transmission; the paper \cite{li2023non} leverages NOMA with quantization modules to enable intelligent multi-user detection {\color{black}(IMUD)}; the work \cite{zhang2024interference} employs semantic differential superposition coding at the transmitter and interference cancellation at the receiver {\color{black}(SeDSIC)} to support both homogeneous and heterogeneous transmission (i.e., semantic information and bit streams); the paper \cite{bo2025deep} extracts and decorrelates both basic and enhanced source features, then superposes them probabilistically for transmission, allowing receivers to decode semantic information commensurate with channel quality; {\color{black}the authors of \cite{yanadaptive} propose semantic adaptive feature extraction (SAFE) network and integrated NOMA for downlink transmission; the work \cite{mengprompt} integrates rate splitting, i.e., dividing information into public (non-orthogonal transmission) and private (orthogonal transmission) components, with generative adversarial networks (GAN)-based interference mitigation.}
However, these methods are still insufficient in terms of interference optimization in NOMA scenarios and cannot be applied to dynamic channel conditions. Additionally, the authors of \cite{tong2024multi} proposes Swin-Transformer-based feature fusion schemes under degraded broadcast channels (DBC) to reduce transmission overhead via joint representations, though the high computational complexity limits scalability in multi-user scenarios. Novel semantic multiple-access schemes have also emerged: the paper \cite{liang2024orthogonal} exploits DeepJSCC’s interference resilience to propose an orthogonal model division approach, while the work \cite{wang2025generative} introduces semantic feature multiple access for multi-user video transmission. {\color{black}The authors of \cite{masemantic} propose a semantic feature domain interference management scheme for multi-user interference channels.}
Despite integrating traditional multiple-access principles with neural network robustness, these solutions lack systematic strategies for interference elimination and resource optimization in multi-user non-orthogonal settings. This paper focuses on uplink multi-user non-orthogonal semantic transmission and designs a low-complexity deep learning framework that can adapt to different channel conditions and further optimize to reduce the impact of interference.%, while enhancing image quality at both pixel and perceptual levels through generative models.
\subsection{Knowledge Distillation and Its Application in Semantic Communication} \label{kdsemcom}
Knowledge distillation, first proposed by Hinton et al. \cite{hinton2015distilling}, aims to transfer the representational capacity of teacher model to student model. The authors of \cite{gou2021knowledge} categorize distilled knowledge into three types: target knowledge, encompassing soft targets from the teacher model typically used in classification tasks; feature knowledge, derived from intermediate layer outputs; and relationship knowledge, capturing inter-layer or inter-sample relationships. 
This framework has been widely used in tasks such as object detection and image super-resolution \cite{wang2024crosskd, he2020fakd}. Recently, knowledge distillation has been adapted to semantic communication. For example, the paper \cite{nguyen2024optimizing} enhances the performance of users with varying computational capabilities in downlink multi-user scenarios, outperforming iterative training methods, while the work \cite{liu2023knowledge} leverages distillation to address multi-user interference and random noise, improving classification performance in text-based semantic systems. However, these studies primarily focus on optimizing performance for specific desired models, neglecting the overall system’s average performance, and have not yet to explore the application of knowledge distillation in multi-user non-orthogonal transmission scenario. Moreover, their reliance on prediction mimicry for model compression risks misalignment between the student network’s optimization objectives and semantic transmission requirements. %Addressing the multi-user interference challenges identified in Section \ref{musemcom}, 
This paper introduces knowledge distillation into uplink multi-user image semantic NOMA transmission, which achieves a comprehensive improvement in the quality of image pixel reconstruction without increasing the inference overhead.
\subsection{Diffusion Model and Its Application in Semantic Communication}
Conventional DeepJSCC frameworks typically aim to minimize pixel-level distortion, such as mean squared error (MSE), as their optimization objective, but they ignore the improvement of image perceptual quality, especially in poor channel environments. In recent years, generative models have achieved groundbreaking progress in computer vision, with diffusion models emerging as a superior alternative to GANs due to their capacity to produce high-quality and semantically coherent images. Within the domain of semantic communication, GANs have been employed to improve image transmission quality. For instance, the paper \cite{erdemir2023generative} models the entire DeepJSCC transmission pipeline as the forward process, leveraging a pretrained StyleGAN2 to solve the inverse problem and improve perceptual quality. In contrast, diffusion models offer distinct advantages in stability, resistance to mode collapse, and generative diversity. Accordingly, the authors of \cite{yang2024diffusion} introduce the diffusion-aided JSCC framework, utilizing a pretrained Stable Diffusion model for the image reconstruction with initial JSCC decoding and multimodal conditions (spatial, textual, and channel state information). {\color{black} The work \cite{wangdiffcom} introduces DiffCom which leverages channel received signal as condition to guide stochastic posterior sampling.} %, enhancing perceptual quality under low-rate and low SNR conditions. 
Similarly, the works of \cite{yilmaz2024high} and \cite{qiao2024latency} deploy pretrained diffusion models at the receiver to boost perceptual quality. %while the paper of \cite{grassucci2023generative} proposes a generative semantic communication framework that harnesses the diffusion models to generate high-quality and semantically identifiable images under high semantic compression and extremely noisy channels. 
Furthermore, the work \cite{wu2024cddm} proposes channel denoising diffusion models (CDDM), which leverage diffusion models to preprocess channel outputs. The paper \cite{zhang2025semantics} presents the semantics-guided diffusion DeepJSCC (SGD-JSCC) scheme, which integrates diffusion models with multi-modal semantic guidance (i.e., text and edge maps) to design a channel denoising model. Inspired by these advancements, this paper innovatively proposes a two-stage image reconstruction framework tailored for semantic NOMA scenarios, deploying pretrained diffusion models at the receiver to achieve superior perceptual quality.

{\color{black}In sum, We identify critical research gaps in these domains and provide a comprehensive comparison with our KDD-SemNOMA framework in Table \ref{tab:comparison_noma_works}.}
\begin{table*}[t!]
	\color{black}
	\captionsetup{font={color={black}}}
	\caption{Comparison with representative multi-user semantic communication methods}
	\centering
	\setlength{\tabcolsep}{2.5pt} %缩小列间距
	\begin{tabular}{lccccccc}
		\toprule
		\textbf{Method} & \textbf{Dataset(s)} & \textbf{Channel(s)} & \textbf{Link direction} & \textbf{Interference Management} & \textbf{Generative Model} &\textbf{Metric(s)}\\
		\midrule
		KDD-SemNOMA  & CIFAR-10, FFHQ-256 & AWGN, Rayleigh & Uplink & Knowledge Distillation & Diffusion Model &PSNR, FID \\ \midrule
		DeepJSCC-NOMA \cite{yilmaz2023distributed}  & CIFAR-10 & AWGN & Uplink & Curriculum Learning & No &PSNR\\
		SemCom\cite{liu2023knowledge} & Europarl & Rayleigh & Uplink & Knowledge Distillation & No & BLEU \\
		NOMASC\cite{li2023non}  & CIFAR-10, MINIST, Europarl & AWGN, Rayleigh & Downlink & IMUD & No & PSNR, BLEU \\
		IS-SNOMA\cite{zhang2024interference} & Cityscapes & Rician & Downlink & SeDSIC & No & PSNR \\
		DeepSCM\cite{bo2025deep} &CIFAR-100 & AGWN & Downlink & LMMSE Decorrelator & No & PSNR\\
		DBC-Aware SC\cite{tong2024multi} & CIFAR-10, STL-10, CelebA & AWGN & Downlink & Semantic Fusion & No & PSNR\\
		SAFE\cite{yanadaptive} &ImageNet &AWGN, Rayleigh & Downlink & Three-stage Training & No & PSNR\\
		SFMA\cite{wang2025generative} &CIFAR-10, SNU-FILM & AWGN & Downlink & User Pairing Algorithm & TAIN & PSNR, LPIPS \\
		SFDMA\cite{masemantic} & MNIST, CelebA & Rayleigh & Bidirectional & Information Bottleneck & No & PSNR\\
		DeepPASIC\cite{mengprompt} & ImageNet & AWGN & Bidirectional & Two-stage Transmission & GAN & PSNR\\
		\bottomrule
	\end{tabular}
	\label{tab:comparison_noma_works}
	\end{table*}

\section{System Model} \label{System Model}
\subsection{Overview}
\begin{figure}[!t]
	\centering
	\includegraphics[width=3.5in]{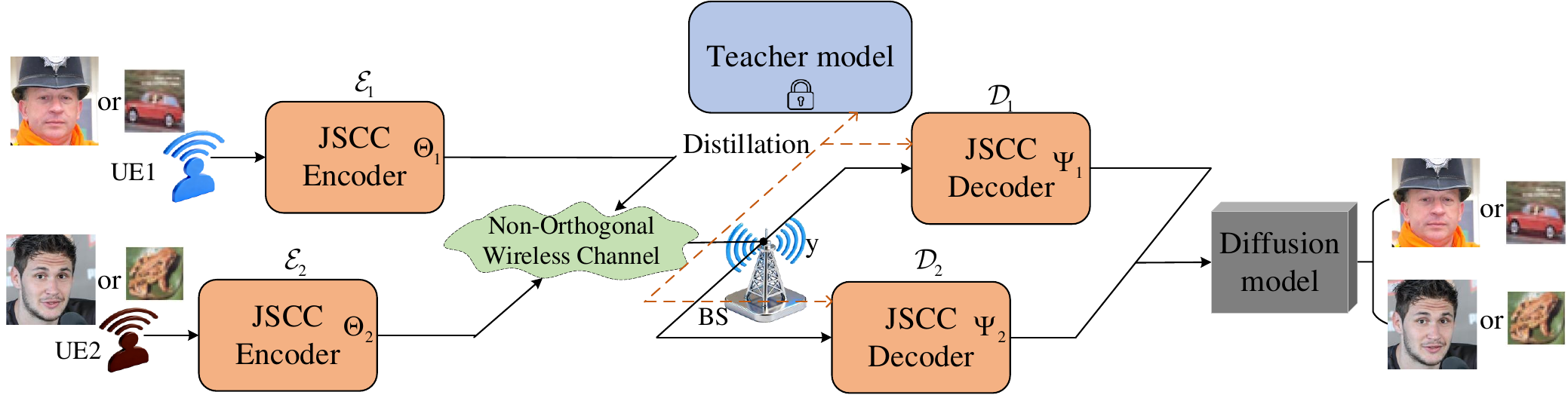}
	\captionsetup{font={footnotesize,{color=black}},justification=raggedright,singlelinecheck=false}
	\caption{\color{black}Overview of the proposed KDD-SemNOMA Framework.}
	\label{Fig0}
	\vspace{-2mm}
\end{figure}

The proposed KDD-SemNOMA framework is illustrated in Figure \ref{Fig0}. Consider an uplink multi-user image semantic communication scenario where multiple user equipment (UE) transmit uplink image semantic information to the base station (BS). Unlike conventional communication systems employing separate source and channel coding, our framework adopts JSCC to directly transmit semantic features. Both UE and BS are configured with single antennas. Each UE transmits an image $\mathbf{x}_{i} \in \mathbb{R}^{C \times H \times W}$, where $i$ denotes the index of UE, $C$, $H$, and $W$ represent the color channels, width, and height of the RGB image, respectively. We denote $m=C\times H\times W$ {\color{black}as the total number of pixels in input image.} The transmitter encodes the source image $\mathbf{x}_{i}$ into compressed semantic features $\mathbf{s}_{i} \in \mathbb{C}^k$ through a JSCC encoder, which can be expressed as
\begin{equation}
	{\mathbf{s}_{i}} = \mathcal{E}_{i}({\mathbf{x}_{i}; \mathbf{\Theta}_{i}}), i \in\{1,2, \cdots, N\},
	\vspace{-1.5mm}
\end{equation}
where $N$ is the total number of UE, $\mathcal{E}_{i}(\cdot)$ denotes the encoder neural network of the $i$-th UE with the trainable parameters $\mathbf{\Theta}_{i}$. The bandwidth compression ratio $\rho$, representing the available channel symbols, is defined as $\rho = \frac{k}{m}$. 
An average transmit power constraint is imposed on $\mathbf{s}_{i}$, such that
\begin{equation}
	\frac{1}{k}\left\|\mathbf{s}_i\right\|_2^2 \leq P_{\text{avg}}. \label{eq:power_constraint}
	\vspace{-1.5mm}
\end{equation} 
%The bandwidth ratio $\rho$, representing the available channel resources, is defined as:
%\begin{equation}
%	\rho = \frac{k}{CHW} \quad \text{(channel symbols/pixel)}.
%\end{equation}
To reduce transmission overhead, the semantic features from multiple users are non-orthogonally superimposed on the same time-frequency resources and transmitted through the channel. The received signal at the BS can be expressed as
\begin{equation}
	\mathbf{y} = \sum_{i=1}^N {h}_i\mathbf{s}_{i} + \mathbf{n},
	\vspace{-1.5mm}
\end{equation}
where $
h_{i} \in \mathbb{C}$ denotes the wireless channel gain between the $i$-th UE and the BS, $\mathbf{n} \in \mathbb{C}^k$ represents the independent and identically distributed (i.i.d.) complex Gaussian noise term with variance $\sigma^2$, i.e., $\mathbf{n} \sim \mathcal{CN}(0, \sigma^2\mathbf{I}_k)$. For AWGN channel, $h_{i}$ is set to $1$. For the Rayleigh fading channel, the channel gain $h_i$ of the $i$-th UE follows $h_i \sim \mathcal{CN}(0, \sigma_h^2) $, where $\sigma_h^2 = 1$ denotes the normalized channel power. We assume $\sigma^2$ and $h_{i}$ are known at both transmitter and receiver. The SNR of the wireless channel is defined as $\gamma=10 \log _{10}\left(\frac{P_{\text{avg}}}{\sigma^2}\right) \mathrm{dB}$.
%\begin{equation}
%	\mathrm{SNR}=10 \log _{10}\left(\frac{P_{\text{avg}}}{\sigma^2}\right) \mathrm{dB}.
%\end{equation}

At the receiver, the JSCC decoder reconstructs the multi-user image based on the superimposed semantic features, i.e., 
\begin{equation}
	{\hat{\mathbf{x}}_i} = \mathcal{D}_{i}(\mathbf{y}; \mathbf{\Psi}_{i}), i \in\{1,2, \cdots, N\}, \label{eq:decoding}
	\vspace{-1.5mm}
\end{equation}
where $\mathcal{D}_{i}(\cdot)$ is the decoder neural network for the \(i\)-th UE with the trainable parameters $\mathbf{\Psi}_{i}$, and \(\hat{\mathbf{x}}_i \in \mathbb{R}^{C \times H \times W}\) is the reconstructed image.

Our framework for multi-user image reconstruction at the receiver comprises two primary stages: (1) multi-user DeepJSCC decoding optimized by knowledge distillation as detailed in Section \ref{KD}, and (2) image refinement guided by diffusion-based generative priors as detailed in Section \ref{Diffusion}. At the first stage, the SemNOMA model is trained using knowledge distillation to effectively decode transmitted signals. At the second stage, we leverage pretrained diffusion model priors to refine low-quality received images, where their visual fidelity and detail have further enhanced.
%\vspace{-2mm}
\subsection{Enhanced AF-Module} \label{EAFModule}
{\color{black}Wireless channels degrade semantic features via additive noise (AWGN channel) and multiplicative fading (Rayleigh channel), which distort feature distributions and impair reconstruction. In NOMA, these effects compound with multi-user interference, severely challenging feature recovery.}

\begin{figure}[!t]
	\centering
	\includegraphics[width=3.6in]{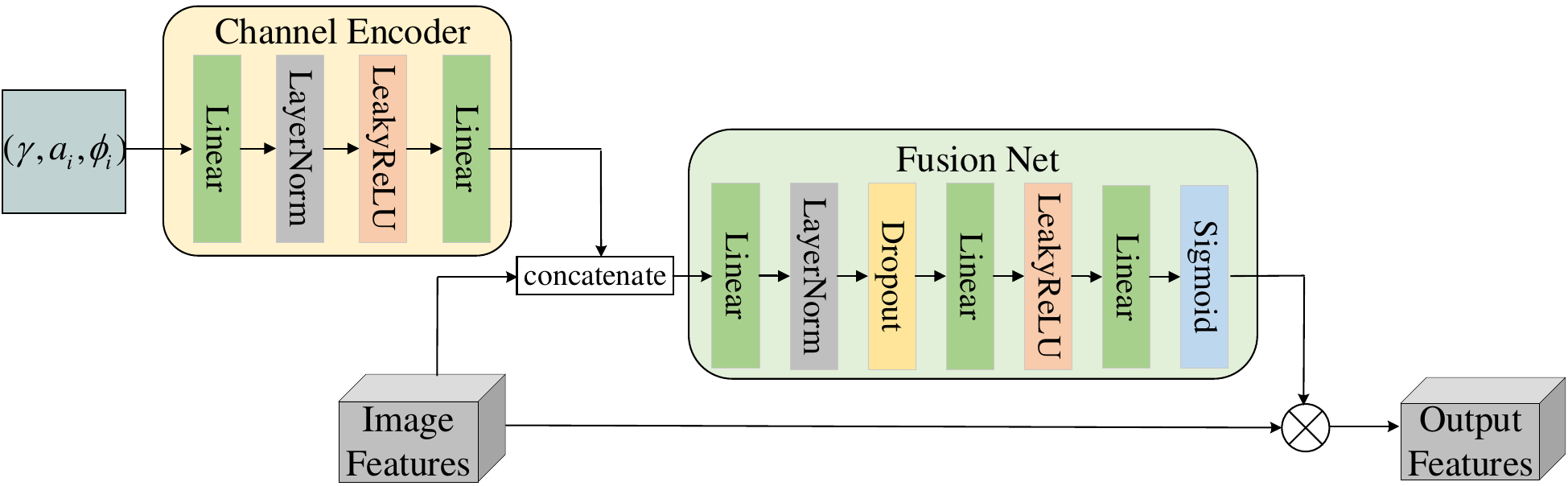}
	\captionsetup{font={footnotesize},justification=raggedright,singlelinecheck=false}
	\caption{Detail architecture of the enhanced AF-Module.}
	\label{Fig2}
	\vspace{-2mm}
\end{figure}

Inspired by \cite{xu2021wireless}, to improve the adaptability and robustness of the SemNOMA framework under varying channel conditions, we propose an enhanced attention feature module (AF-Module), as shown in Fig.~\ref{Fig2}. The module dynamically modulates image semantic features using channel side information, including SNR and Rayleigh fading parameters, enabling robust performance across diverse channel environments.

Let the input semantic feature map be $\mathbf{f}_s \in \mathbb{R}^{B \times C' \times H' \times W'}$, where $B$, $C'$, $H'$, and $W'$ denote batch size, number of channels, height, and width, respectively. The channel side information is represented by $\mathbf{s} \in \mathbb{R}^{B \times D}$, with $D=3$ corresponding to the SNR $\gamma \in [0, 20]$ dB and Rayleigh fading parameters of the $i$-th UE (amplitude $a_i$ and phase $\phi_i$). During training, $\gamma$ is uniformly sampled from $[0, 20]$ dB, and the Rayleigh parameters $(a_i, \phi_i)$ are randomly generated to simulate diverse channel conditions. %This enhances the model's adaptability and reduces the impact of channel on the semantic features.
The enhanced AF-Module processes $\mathbf{f}_s$ and $\mathbf{s}$ through the following steps:
\begin{enumerate}
	\item \textbf{Spatial context extraction}: The spatial context feature $\mathbf{f}_{\text{ctx}} \in \mathbb{R}^{B \times C'}$ is computed by averaging $\mathbf{f}_s$ over spatial dimensions
	\begin{equation}
		\mathbf{f}_{\text{ctx}} = \frac{1}{H' \cdot W'} \sum_{h=1}^{H'} \sum_{w=1}^{W'} \mathbf{f}_s[b, c, h, w].
		\vspace{-2mm}
	\end{equation}
	\item \textbf{Channel feature encoding}: The side information $\mathbf{s}$ is encoded into a channel feature $\mathbf{f}_c \in \mathbb{R}^{B \times C'}$ via a channel encoder $\mathcal{F}_c$
	%\begin{equation}
	%	\mathbf{f}_c = \text{LeakyReLU} \big( \text{LayerNorm} ( \mathbf{W}_1 \mathbf{s} + \mathbf{b}_1 ) \big) \mathbf{W}_2 + \mathbf{b}_2,
	%\end{equation}
	\begin{equation}
		\mathbf{f}_c = \mathcal{F}_c(\mathbf{s}), 
		\vspace{-2mm}
	\end{equation}
	where $\mathcal{F}_c$ consists of two linear layers with LayerNorm and Leaky rectified linear unit (LeakyReLU) activation.
	%where $\mathbf{W}_1 \in \mathbb{R}^{128 \times D}$, $\mathbf{W}_2 \in \mathbb{R}^{C \times 128}$ are weight matrices, and $\mathbf{b}_1$, $\mathbf{b}_2$ are biases.
	\item \textbf{Feature fusion}: The spatial context and channel features are concatenated to form $\mathbf{f}_f \in \mathbb{R}^{B \times 2C'}$
	\begin{equation}
		\mathbf{f}_f = \operatorname{Concat}(\mathbf{f}_{\text{ctx}}, \mathbf{f}_c),
		\vspace{-2mm}
	\end{equation}
	where $\operatorname{Concat}(\cdot)$ operates along the channel dimension.
	\item \textbf{Adaptive mask generation}: An adaptive mask $\mathbf{m} \in \mathbb{R}^{B \times C' \times 1 \times 1}$ is generated via a fusion network $\mathcal{F}_m$
	%\begin{equation}
	%	\begin{aligned}
	%		\mathbf{m} &= \sigma \big( \text{LeakyReLU} \big( \mathbf{W}_5 \cdot \text{LeakyReLU} \big( \mathbf{W}_4 \cdot \text{Dropout} \\
	%		&\quad ( \mathbf{W}_3 \mathbf{f}_f + \mathbf{b}_3 ) + \mathbf{b}_4 \big) + \mathbf{b}_5 \big) \big),
	%	\end{aligned}
	%\end{equation}
	\begin{equation}
		\mathbf{m} = \mathcal{F}_m(\mathbf{f}_f), 
		\vspace{-1.5mm} 
	\end{equation}
	%where $\mathbf{W}_3 \in \mathbb{R}^{4C \times 2C}$, $\mathbf{W}_4 \in \mathbb{R}^{4C \times 4C}$, $\mathbf{W}_5 \in \mathbb{R}^{C \times 4C}$ are weight matrices, $\mathbf{b}_3$, $\mathbf{b}_4$, $\mathbf{b}_5$ are biases, and $\text{Dropout}(0.1)$ applies a 10\% dropout rate.
	where $\mathcal{F}_m$ comprises three linear layers with LeakyReLU, $10\%$ dropout, and sigmoid function.

	\item \textbf{Feature modulation}: The semantic features $\mathbf{f}_s$ are modulated using the mask:
	\begin{equation}
		\mathbf{f}_{\text{out}} = \mathbf{f}_s \odot \operatorname{Broadcast}(\mathbf{m}),
		\vspace{-1.5mm}
	\end{equation}
	where $\operatorname{Broadcast}(\cdot)$ aligns $\mathbf{m} \in \mathbb{R}^{B \times C' \times 1 \times 1}$ with the spatial dimensions of $\mathbf{f}_s \in \mathbb{R}^{B \times C' \times H' \times W'}$, $\odot$ denotes element-wise multiplication.
\end{enumerate}
{\color{black}In summary, the enhanced AF-Module leverages channel state information to dynamically recalibrate features. It produces channel-aware attention masks that selectively emphasize robust components while suppressing noise-corrupted ones, enabling adaptive compensation for channel distortions.}

%\vspace{-4mm}
\subsection{SemNOMA Network Architecture} \label{SemNOMA_arch}
% 跨双栏的 Figure
\begin{figure*}[!t]
	\centering
	\includegraphics[width=0.88\linewidth]{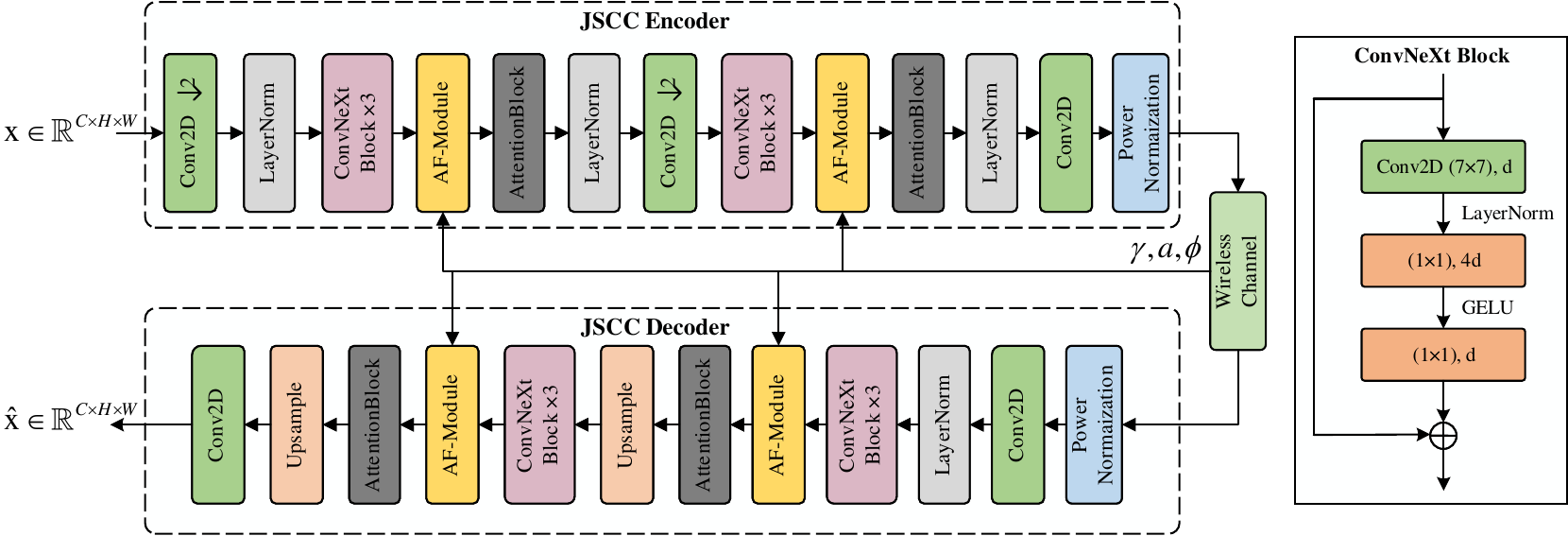}
	\captionsetup{font={footnotesize},justification=raggedright,singlelinecheck=false}
	\caption{Architecture of the ConvNeXt-based DeepJSCC Network, where Conv2D $\downarrow 2$ represents 2-dimensional convolution with stride=2 for downsampling.}
	\label{Fig1}
	\vspace{-2mm}
\end{figure*}
Fig. \ref{Fig1} illustrates the proposed DeepJSCC neural network, designed for efficient compression and reconstruction of image source over noisy channels. The architecture is based on the modified ConvNeXt framework\cite{liu2022convnet}, where downsampling, upsampling, and attention modules are incorporated.

%Fig. \ref{Fig1} shows the DeepJSCC neural network structure we proposed. The encoder and decoder are designed to efficiently compress and reconstruct multi-user image data over the noisy channel, leveraging a modified ConvNeXt framework \cite{liu2022convnet} with integrated downsampling, upsampling, and attention modules. 

The ConvNeXt block, which modernizes ResNet with Vision Transformer-inspired designs, achieving state-of-the-art on ImageNet classification task and outperforming Swin Transformer on COCO detection challenges while maintaining the maturity and simplicity of convolutional neural networks \cite{liu2022convnet}. As shown in Fig. \ref{Fig1}, each ConvNeXt block processes input features of dimension $C'$ through a sequence of operations: a depthwise convolution with a $7 \times 7$ kernel, followed by a LayerNorm applied in a channels-last format. This is succeeded by a pointwise convolution implemented as a linear layer expanding the dimension to $4C'$, a Gaussian error linear unit (GELU) activation for nonlinearity, and another linear layer reducing it back to $C'$. A residual connection with stochastic depth enhances training stability, balancing efficiency and expressive power for feature extraction in DeepJSCC.

The encoder processes RGB images and normalizes them to the $[0, 1]$ range by dividing pixel values by 255. It employs a stem layer for initial downsampling, followed by LayerNorm and additional downsampling stages. Multiple ConvNeXt blocks, with $N_e$ blocks per stage, process features, as detailed in Table~\ref{tab:encoder_decoder_cifar10} for 32$\times$32 resolution and Table~\ref{tab:encoder_decoder_ffhq256} for 256$\times$256 resolution. Enhanced AF-Module, as introduced in Section \ref{EAFModule}, improve the robustness to channel noise by conditioning on SNR and Rayleigh fading parameters (i.e., $\gamma, a, \phi$). The final $1 \times 1$ convolution produces $M$ channels, normalized to meet the power constraint (i.e., Eq.~\ref{eq:power_constraint}), and reshaped into $k$ complex-valued channel inputs.

The decoder reconstructs image from the received multi-user superimposed semantic features. The $k$ complex-valued samples are converted to $2k$ real-valued channels, processed by a $1 \times 1$ convolution and LayerNorm. Progressive upsampling stages then restore spatial resolution (e.g., $32 \times 32$ or $256 \times 256$), each followed by $N_d$ ConvNeXt blocks with parametric rectified linear unit (PReLU) activations and AF-Modules for feature refinement, as specified in Table~\ref{tab:encoder_decoder_cifar10} and Table~\ref{tab:encoder_decoder_ffhq256}. A final $1 \times 1$ convolution ($K=3$) with sigmoid activation produces a $C=3$ output, which is then denormalized to range $[0, 255]$.

Building upon the designed DeepJSCC architecture, we constructed a network for multi-user semantic non-orthogonal transmission scenarios, called SemNOMA. Specifically, at the transmitter, we employ a parameter-shared encoder to reduce computational overhead. Following the approach proposed in \cite{yilmaz2023distributed}, we embed user-specific embeddings \(\mathbf{r}_i \in \mathbb{R}^{1 \times H \times W}\), initialized from \(\mathcal{N}(0, 1)\), to enable the model to distinguish users. For each UE $i$, the input image \(\mathbf{x}_i \in \mathbb{R}^{C \times H \times W}\) is concatenated with $\mathbf{r}_i$ along the channel dimension, forming the encoder input \(\mathbf{z}_i \in \mathbb{R}^{(C+1) \times H \times W}\), expressed as
\begin{equation}
	\mathbf{z}_i = \operatorname{Concat}(\mathbf{x}_i, \mathbf{r}_i),
	\vspace{-2mm}
\end{equation}
where \(\operatorname{Concat}(\cdot)\) denotes channel-wise concatenation, preserving spatial dimensions \(H \times W\). At the receiver, multiple decoders are utilized to reconstruct user-specific images, enabling tailored reconstruction for each UE.
%\vspace{-2mm}
\section{Knowledge Distillation Optimization for Semantic NOMA (KD-SemNOMA)} \label{KD}
% 跨双栏的 Figure
\begin{figure*}[!t]
	\centering
	\includegraphics[width=0.90\linewidth]{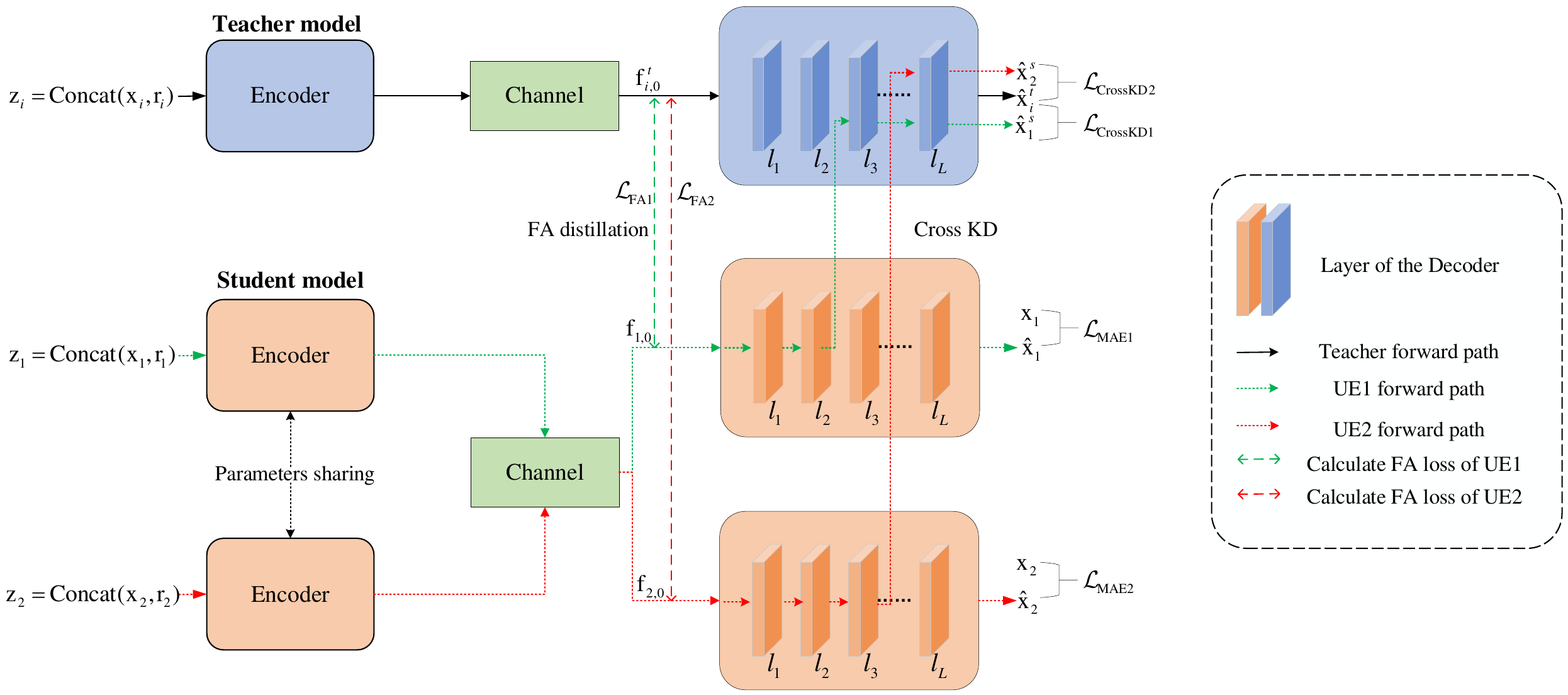}
	\captionsetup{font={footnotesize}}
	\caption{The overall framework of KD-SemNOMA. The teacher model employs orthogonal channel for image semantic features transmission, while the student model utilizes non-orthogonal channel. $l_j$ denotes the $j$-th layer of the decoder. $\hat{x}_i$ denotes the student model output of the $i$-th UE, $\mathbf{f}_{i,0}^t$ denotes the input feature of the teacher decoder, $\mathcal{L}_{\text{MAEi}}$ denotes the MAE loss of the $i$-th UE, $\mathcal{L}_{\text{CrossKDi}}$ denotes the CrossKD loss of the $i$-th UE, $\mathcal{L}_{\text{FAi}}$ denotes the FA loss of the $i$-th UE.}
	\label{Fig3}
	\vspace{-2mm}
\end{figure*}

\begin{algorithm}[!t]
	\caption{Training and inference procedure for SemNOMA with knowledge distillation}
	\label{alg:kd}
	\begin{algorithmic}[1]
		%\Require Multi-user image dataset \( \mathbf{x} = \{ \mathbf{x}_1, \dots, \mathbf{x}_N \} \), hyperparameters \( \lambda_1, \lambda_2, \lambda_3 \)
		{\color{black} \Require Multi-user image training dataset \( \mathbf{x_{train}} = \{ \mathbf{x}_1^{train}, \dots, \mathbf{x}_N^{train} \} \), test dataset \(\mathbf{x_{test}} = \{ \mathbf{x}_1^{test}, \dots, \mathbf{x}_N^{test} \} \), hyperparameters \( \lambda_1, \lambda_2, \lambda_3 \)}
		\Ensure Reconstructed multi-user images \( \hat{\mathbf{x}} = \{ \hat{\mathbf{x}}_1, \dots, \hat{\mathbf{x}}_N \} \)
		
		% ====== Training Phase ======
		\Statex \textbf{Training Phase (Teacher \& Student)}
		\State \textit{Step 1: Pre-train Teacher Model}
		\State Initialize teacher model \( \mathcal{F}_{\text{teacher}} \) parameters
		\State Transmit features orthogonally according to Eq.~\ref{eq:orthogonal_transmission}
		\State Optimize teacher model via MAE loss (Eq.\ref{eq:teacher_loss}): 
		\State Freeze \( \mathcal{F}_{\text{teacher}} \) weights after convergence
		
		\State \textit{Step 2: Train Student Model}
		\State Initialize student model \( \mathcal{F}_{\text{student}} \) with teacher model weights
		%\For{each batch \( \mathbf{x}_{\text{batch}} \subseteq \mathbf{x} \)}
		\color{black}\For{each batch \( \mathbf{x}_{\text{batch}} \subseteq \mathbf{x_{train}} \)}
		\color{black}\State Extract teacher  model features \( \mathbf{f}^t \) and student model features \( \mathbf{f}^s \)
		\State Compute affinity matrices \( \mathbf{A}_i^t, \mathbf{A}_i^s \) (Eq.\ref{eq:affinity_matrix})
		\State Calculate feature affinity loss \( \mathcal{L}_{\mathrm{FA}} \) (Eq.\ref{eq:feature_loss})
		\State Generate cross-head predictions \( \hat{\mathbf{x}}^s \)
		\State Compute CrossKD loss \( \mathcal{L}_{\text{CrossKD}} \) (Eq.\ref{eq:crosskd_loss})
		\State Calculate restoration loss \( \mathcal{L}_{\text{MAE}} \) (Eq.\ref{eq:restoration_loss})
		\State Total loss: \( \mathcal{L}_S = \lambda_1\mathcal{L}_{\text{MAE}} + \lambda_2\mathcal{L}_{\mathrm{FA}} + \lambda_3\mathcal{L}_{\text{CrossKD}} \) %(Eq.\ref{eq:total_loss})
		\State Update \( \mathcal{F}_{\text{student}} \) via \( \nabla \mathcal{L}_S \)
		\EndFor
		
		% ====== Inference Phase ======
		\Statex \textbf{Inference Phase (Student Only)}
		\State Input transmisstion images \( \mathbf{x}_{\text{test}} \)
		\State Forward pass through \( \mathcal{F}_{\text{student}} \): 
		\Statex \hspace{1em} \( \hat{\mathbf{x}} = \mathcal{F}_{\text{student}}( \mathbf{x}_{\text{test}} ) \)
		\State \Return \( \hat{\mathbf{x}} \)
	\end{algorithmic}
	\vspace*{-1mm}  % 算法后压缩
\end{algorithm}
The pipeline of our proposed knowledge distillation optimization strategy for SemNOMA (KD-SemNOMA) is illustrated in Fig. \ref{Fig3}. Multi-user images are processed through dual branches: the teacher model and the student model, both sharing identical network architectures. Orthogonal transmission of multi-user image semantic features is more conducive to training compared to non-orthogonal transmission, as orthogonal features are solely affected by the wireless channel noise, without interference from other users. Inspired by knowledge distillation scheme\cite{hinton2015distilling}, we designate the E2E network with orthogonal transmission as the teacher model and the SemNOMA model as the student model. By effectively transferring user-specific semantic orthogonal knowledge from the teacher model to the student model, the decoding performance of the student SemNOMA model is further enhanced. The training process comprises two stages: pre-training the teacher model, followed by training the student model while freezing the weight parameters of the teacher model. During the inference phase, the teacher model branch is discarded, and only the student network is deployed, ensuring that our approach incurs no additional computational overhead for inference. The complete training and inference workflow of the knowledge distillation scheme is presented in Algorithm \ref{alg:kd}.
%\vspace{-3mm}
\subsection{Teacher Model Training}
To obtain interference-free feature vectors as ``knowledge" for guiding the non-orthogonal transmission network training, we employ an orthogonal transmission scheme as the teacher model. In this configuration, multi-user image semantic features are transmitted orthogonally through the channel. For the $i$-th UE, semantic features $\mathbf{s}_i$ are transmitted through the channel as
\begin{equation}
	\mathbf{y}_i = h_i \mathbf{s}_i + \mathbf{n}, \label{eq:orthogonal_transmission}
	\vspace{-2mm}
\end{equation}
At the receiver, the decoder reconstructs the images with higher quality due to the absence of inter-user interference. We adopt the mean absolute error (MAE) as the loss function $\mathcal{L}_T$ to optimize the teacher model since its efficiency has been widely recognized in such tasks \cite{zou2020deep}, i.e.,
\begin{equation}
	\mathcal{L}_T = \sum_{i=1}^N \left\| \mathcal{F}_{\text{teacher}}(\mathbf{x}_i) - \mathbf{x}_i \right\|_1, \label{eq:teacher_loss}
	\vspace{-2mm}
\end{equation}
where $\mathcal{F}_{\text{teacher}}(\cdot)$ denotes the teacher network’s mapping function, and $\mathbf{x}_i$ is the input image for the $i$-th UE.

%\vspace{-3mm}
\subsection{Student Model Training}
During the training phase of the student SemNOMA model, we initialize the weights of the student model with the parameters from the pre-trained teacher model. This transfers the reconstruction capability of the teacher model to the student model and provides a good starting point for optimization. To effectively transfer knowledge from the teacher model to the student model, we employ both feature distillation and cross head prediction distillation to enhance the learning capacity of the student model. 

For feature distillation, the objective is to align the latent semantic representations of the student and teacher models as closely as possible, thereby enabling the student model to reconstruct interference-free multi-user images. We select feature affinity (FA) knowledge distillation \cite{he2020fakd} as the feature distillation loss function to extract high-dimensional feature semantic information. Specifically, for the $i$-th UE, given a batch of image feature maps $\mathbf{f}_i \in \mathbb{R}^{B \times C' \times H' \times W'}$, we first reshape it into the three-dimensional tensor $\mathbf{f}_i^{'} \in \mathbb{R}^{B \times C' \times H'W'}$ and normalize the feature map to $\tilde{\mathbf{f}}_i[b,:, w]=\mathbf{f}_i^{\prime}[b,:, w] /\left\|\mathbf{f}_i^{\prime}[b,:, w]\right\|_2^2$. We then compute the spatial affinity similarity matrix $\mathbf{A}_i\in \mathbb{R}^{B \times H'W' \times H'W'}$ by inner product, i.e., 
\begin{equation}
	\mathbf{A}_i=\tilde{\mathbf{f}}_i^T \cdot \tilde{\mathbf{f}}_i. \label{eq:affinity_matrix}
	\vspace{-1.5mm}
\end{equation}
The student model is encouraged to produce similar affinity matrices with teacher model and the feature affinity-based distillation loss $\mathcal{L}_{\mathrm{FA}}$ can be formulated as
\begin{equation}
	\mathcal{L}_{\mathrm{FA}} = \sum_{i=1}^N \frac{1}{|\mathbf{A}_i|_c} \sum_{j=1}^{L} \left\| \mathbf{A}_{i,j}^s - \mathbf{A}_{i,j}^t \right\|_1. \label{eq:feature_loss}
	\vspace{-1.5mm}
\end{equation}
Note that we utilize the MAE loss to optimize the similarity of the spatial feature affinity matrices, where $\mathbf{A}_{i,j}^t$ and $\mathbf{A}_{i,j}^s$ denote the affinity matrices for the $i$-th UE, extracted from the $j$-th layer feature maps of the teacher and student models respectively. $L$ is the number of layers we choose to extract features. $|\mathbf{A}_i|_c$ denotes the number of elements in the affinity matrix. In particular, we adopt the input features of the teacher and student models' decoders to calculate the feature distillation loss, respectively.

For prediction distillation, instead of directly minimizing the discrepancy between the predictions of teacher model and the student model, we adopt the cross-head knowledge distillation (CrossKD) \cite{wang2024crosskd} to alleviate the conflict between the supervised and distillation targets. Specifically, CrossKD transfers the intermediate semantic features from the student model’s head to that of the teacher model to produce the cross head predictions for distillation, which implicitly builds the connection between the heads of the teacher-student pair to improve the distillation efficiency. In particular, we implement CrossKD on the decoder. For the $i$-th UE, let $\mathbf{f}_{i,j}, j \in\{1,2, \cdots, L\}$ denoted the feature maps produced by the $j$-th decoder layer $l_j$, with $\mathbf{f}_{i,0}$ being the input feature map of the decoder. The restored images of the teacher and the student model of the $i$-th UE can be represented as $\hat{\mathbf{x}}_i^t \in \mathbb{R}^{B \times C \times H \times W}$ and $\hat{\mathbf{x}}_i^s \in \mathbb{R}^{B \times C \times H \times W}$, respectively. Besides the original restoration from the teacher and the student, CrossKD additionally delivers the student model's intermediate features $\mathbf{f}_{i,j}, j \in\{1,2, \cdots, L\}$ to the $(j+1)$-th decoder layer $l_{j+1}$ of the teacher model, resulting in the cross-head restoration $\hat{{\mathbf{x}}}^s_i \in \mathbb{R}^{B \times C \times H \times W}$. Similar to the feature distillation loss function, we use MAE loss to optimize the image restoration result. The CrossKD objective $\mathcal{L}_{\text{CrossKD}}$ is described as follows
\begin{equation}
	\mathcal{L}_{\text{CrossKD}} = \sum_{i=1}^N \left\| \hat{\mathbf{x}}_i^s - \hat{\mathbf{x}}_i^t \right\|_1. \label{eq:crosskd_loss}
	\vspace{-1.5mm}
\end{equation}
Furthermore, the MAE loss $\mathcal{L}_{\text{MAE}}$ between the ground truth and the output of the student model can be denoted as
\begin{equation}
	\mathcal{L}_{\text{MAE}} = \sum_{i=1}^N \left\| \mathcal{F}_{\text{student}}(\mathbf{x}_i) - \mathbf{x}_i \right\|_1, \label{eq:restoration_loss}
	\vspace{-1.5mm}
\end{equation}
where $\mathcal{F}_{\text{student}}(\cdot)$ denotes the student network’s mapping function. The total loss $\mathcal{L}_S$ for training the student model can be formulated as the weighted sum of the image restoration loss and the distillation loss, i.e., 
%\begin{equation}
%	\mathcal{L} = \sum_{i=1}^N \left\| f_{\text{student}}(\mathbf{x}_i) - \mathbf{x}_i \right\|_1, \label{eq:restoration_loss}
%\end{equation}
%\begin{equation}
%	\mathcal{L}_S = \lambda_1 \mathcal{L} + \lambda_2 \mathcal{L}_{AD} + \lambda_3 \mathcal{L}_{\text{CrossKD}}, \label{eq:total_loss}
%\end{equation}
\begin{align}
	\mathcal{L}_S &= \lambda_1 \mathcal{L}_{\text{MAE}} + \lambda_2 \mathcal{L}_{\mathrm{FA}} + \lambda_3 \mathcal{L}_{\text{CrossKD}}, \label{eq:total_loss}
	\vspace{-1.5mm}
\end{align}
where $\lambda_1$, $\lambda_2$, $\lambda_3$ govern the trade-off among different aspects of the loss. 
{\color{black}
\subsection{Discussion}
In summary, our KD-SemNOMA framework establishes a comprehensive teacher-student interaction protocol that enables effective knowledge transfer across multiple levels that effectively bridges the performance gap between orthogonal and non-orthogonal transmission systems. The key aspects of our approach include:

\begin{enumerate}
	\item \textbf{Multi-level Knowledge Transfer:} During training, we implement knowledge distillation at both feature-level (through Feature Affinity loss) and prediction-level (through CrossKD loss), ensuring comprehensive guidance from the teacher model to the student model.
	
	\item \textbf{Efficient Training Protocol:} The progressive training strategy with teacher model initialization provides a strong foundation, while the adaptive loss balancing mechanism ensures stable optimization throughout the training process.
	
	\item \textbf{Computational Efficiency:} During inference, only the lightweight student model is deployed with no teacher-student interaction, ensuring no additional computational overhead compared to baseline SemNOMA system.
\end{enumerate}

The performance improvement achieved by our method is particularly significant in challenging low-SNR scenarios, where the knowledge distillation effectively mitigates the impact of both channel noise and multi-user interference. While perfect knowledge transfer cannot be guaranteed due to the fundamental constraints of non-orthogonal transmission, our experimental results in Section \ref{experiment of KD-SemNOMA} demonstrate that the student model consistently outperforms baseline methods without knowledge distillation.
}
\section{Diffusion Model-Based Image Refinement} \label{Diffusion}
\begin{figure}[!t]
	\centering
	\includegraphics[width=3.5in]{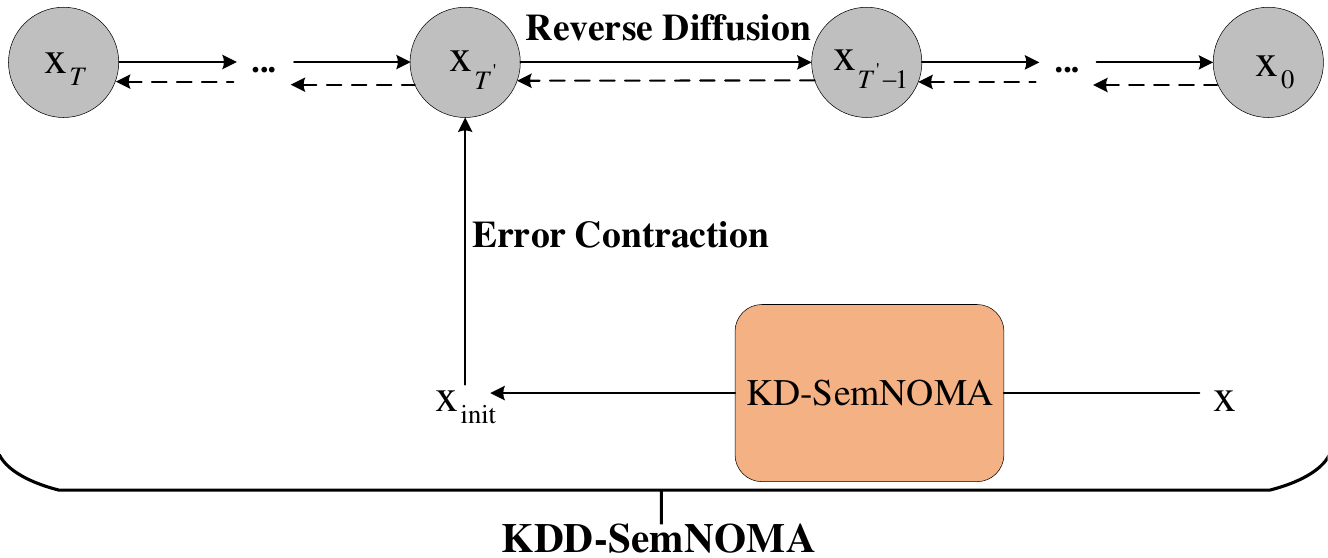}
	\captionsetup{font={footnotesize}}
	\caption{Block diagram of the image refinement based on the pre-trained diffusion model (KDD-SemNOMA).}
	\label{Fig4}
	\vspace{-2mm}
\end{figure}
At the first stage, the proposed KD-SemNOMA framework, trained via knowledge distillation to leverage the semantic knowledge of the teacher model, recovers multi-user images optimized for pixel-level accuracy using MAE. However, this approach often yields suboptimal perceptual quality, particularly under poor channel conditions (e.g., low SNR). To address this, inspired by \cite{yue2024difface}, we further propose a second-stage diffusion model based image restoration method, called KDD-SemNOMA. This leverages the error contraction property of diffusion process and the image priors encapsulated in the pre-trained diffusion model to enhance semantic and perceptual quality of the reconstruction images.
%\vspace{-3mm}
\subsection{Preliminary}
Diffusion model comprises a forward (diffusion) process and a reverse process \cite{ho2020denoising}. Given a data sample $\mathbf{x}_0$ with the probability distribution $q\left(\mathbf{x}_0\right)$, the forward process progressively adds Gaussian noise through $T$ steps according to a pre-defined or learned noise variance schedule $\{\beta_t\}_{t=1}^T$ where $\beta_t \in (0,1)$. This process is defined as a Markov chain with the conditional probability distribution, i.e.,
\begin{equation}
	q\left(\mathbf{x}_t \mid \mathbf{x}_{{t}-{1}}\right)=\mathcal{N}\left(\mathbf{x}_t ; \sqrt{1-\beta_t} \mathbf{x}_{t-1}, \beta_t \mathbf{I}_m\right),
	\vspace{-1.5mm}
\end{equation}
where $t \in\{1,2, \cdots, T\}$. A well-designed variance schedule theoretically guarantees that $q\left(\mathbf{x}_T\right)$ converges to the simple Gaussian distribution. Remarkably, the marginal distribution at arbitrary timestep $t$ has the following analytical form, i.e.,
\begin{equation}
	q(\mathbf{x}_t \mid \mathbf{x}_0) = \mathcal{N}(\mathbf{x}_t; \sqrt{\alpha_t} \mathbf{x}_0, (1 - \alpha_t) \mathbf{I}_m), \label{eq:marginal_diffusion}
	\vspace{-1.5mm}
\end{equation}
where $\alpha_t=\prod_{l=1}^t\left(1-\beta_l\right)$.

The reverse process learns a transition kernel from \(\mathbf{x}_t\) to \(\mathbf{x}_{t-1}\) which is defined as the following Gaussian distribution, i.e.,
\begin{equation}
	p_\theta(\mathbf{x}_{t-1} \mid \mathbf{x}_t) = \mathcal{N}(\mathbf{x}_{t-1}; \boldsymbol{\mu}_\theta(\mathbf{x}_t, t), \sigma_t^2 \mathbf{I}_m), \label{eq:reverse_process}
	\vspace{-1.5mm}
\end{equation}
where \(\theta\) denotes learnable parameters. %With such a learned transition kernel, we can approximate the data distribution $q\left(\mathbf{x}_0\right)$ via the following marginal distribution
%\begin{equation}
%	p_\theta(\mathbf{x}_0) = \int p(\mathbf{x}_T) \prod_{t=1}^T p_\theta(\mathbf{x}_{t-1} \mid \mathbf{x}_t) \, d\mathbf{x}_{1:T}, \label{eq:data_distribution}
%\end{equation}
%where $p\left(\mathbf{x}_T\right)=\mathcal{N}\left(\mathbf{x}_T ; \mathbf{0}, \mathbf{I}_m\right)$.

%In the denoising diffusion probabilistic model (DDPM) \cite{ho2020denoising}, the reverse posterior $p_\theta\left(\mathbf{x}_{t-1} \mid \mathbf{x}_t\right)$ is reparametrized as follows
%\begin{equation}
%	\begin{gathered}
%		\boldsymbol{\mu}_\theta\left(\mathbf{x}_t, t\right)=\sqrt{\frac{\alpha_{t-1}}{\alpha_t}}\left(\mathbf{x}_t-\frac{\beta_t}{\sqrt{1-\alpha_t}} \boldsymbol{\varepsilon}_\theta\left(\mathbf{x}_t, t\right)\right), \\
%		\sigma_t^2=\frac{1-\alpha_{t-1}}{1-\alpha_t} \beta_t,
%	\end{gathered}
%\end{equation}
%\begin{equation}
%	\begin{aligned}
%		\boldsymbol{\mu}_\theta(\mathbf{x}_t, t) &= \sqrt{\frac{\alpha_{t-1}}{\alpha_t}} \left( \mathbf{x}_t - \frac{\beta_t}{\sqrt{1 - \alpha_t}} \boldsymbol{\varepsilon}_\theta(\mathbf{x}_t, t) \right), \\
%		\sigma_t^2 &= \frac{1 - \alpha_{t-1}}{1 - \alpha_t} \beta_t,
%	\end{aligned}
%	\vspace{-1.5mm}
%\end{equation}
%where $\boldsymbol{\varepsilon}_\theta\left(\mathbf{x}_t, t\right)$ is a designed neural network to predict the noise contained in $\mathbf{x}_t$. 
The denoising diffusion implicit model (DDIM) \cite{songdenoising} inference framework generalizes the denoising diffusion probabilistic model (DDPM) \cite{ho2020denoising} to an non-Markov process, the reverse posterior is expressed as
%\begin{equation}
%	\begin{gathered}
%		\boldsymbol{\mu}_\theta\left(\mathbf{x}_t, t\right)=\sqrt{\alpha_{t-1}} \hat{\mathbf{x}}_0^{(t)}+\sqrt{1-\alpha_{t-1}-\sigma_t^2} \boldsymbol{\varepsilon}_\theta\left(\mathbf{x}_t, t\right), \\
%		\sigma_t^2=\eta \cdot \frac{1-\alpha_{t-1}}{1-\alpha_t} \beta_t, \eta \in[0,1],
%	\end{gathered} \label{eq:ddim_posterior}
%\end{equation}
\begin{equation}
	\begin{aligned}
		\boldsymbol{\mu}_\theta(\mathbf{x}_t, t) &= \sqrt{\alpha_{t-1}} \hat{\mathbf{x}}_0^{(t)} + \sqrt{1 - \alpha_{t-1} - \sigma_t^2} \boldsymbol{\varepsilon}_\theta(\mathbf{x}_t, t), \\
		\sigma_t^2 &= \eta \cdot \frac{1 - \alpha_{t-1}}{1 - \alpha_t} \beta_t, \quad \eta \in [0,1],
	\end{aligned}
	\label{eq:ddim_posterior}
	\vspace{-1.5mm}
\end{equation}
where
\begin{equation}
	\hat{\mathbf{x}}_0^{(t)}=\frac{\mathbf{x}_t-\sqrt{1-\alpha_t} \boldsymbol{\varepsilon}_\theta\left(\mathbf{x}_t, t\right)}{\sqrt{\alpha_t}} .
	\vspace{-1.5mm}
\end{equation}
The hyper-parameter $\eta$ controls the extent of randomness during inference. 
%Specifically, when $\eta=0$, the DDIM reverse process is deterministic, while for $\eta=1$ it is equivalent to the DDPM model.

The marginal distribution of Eq.~\ref{eq:marginal_diffusion} reveals that the initial state $\mathbf{x}_0$ is contracted by a scaling factor $\sqrt{\alpha_t}$ after transitioning to timestep $t$. As demonstrated by recent works \cite{yue2024difface}, when $\mathbf{x}_0$ is interpreted as a coarse estimated image, this marginal distribution inherently compresses the prediction error associated with $\mathbf{x}_0$. Motivated by this insight, as shown in Fig. \ref{Fig4}, we propose to treat the image estimated by KD-SemNOMA as the initial estimate $\mathbf{x}_{\text{init}}$. By leveraging the error contraction property of diffusion process, we first apply forward diffusion to $\mathbf{x}_{\text{init}}$ to obtain $\mathbf{x}_{T'}$, then employ the reverse diffusion process to iteratively denoise $\mathbf{x}_{T'}$, thereby enhancing the image perceptual quality.
\subsection{Diffusion-Based Image Refinement}
In our proposed KDD-SemNOMA scheme, we first leverage the proposed KD-SemNOMA model (as depicted in Section \ref{KD}) to reconstruct images $\mathbf{x}_{\text{init}} = \hat{\mathbf{x}}$ from multi-user non-orthogonal transmissions (Eq.~\ref{eq:decoding}). We denote the real high-quality images as $\mathbf{x}$ and the first-stage reconstructed images from the KD-SemNOMA model as $\mathbf{x}_{\text{init}}$. Due to compression losses, channel noise and multi-user interference inherent in the semantic NOMA scenario, $\mathbf{x}_{\text{init}}$ deviates from $\mathbf{x}$, introducing reconstruction errors. To refine $\mathbf{x}_{\text{init}}$ into a high-fidelity approximation of $\mathbf{x}$, we propose designing a posterior distribution $p(\mathbf{x}\mid\mathbf{x}_{\text{init}})$, inspired by the diffusion-based image enhancement strategy in DifFace \cite{yue2024difface}.

The refinement process consists of two primary stages, as depicted below:
\begin{equation}
	\underbrace{\mathbf{x} \xrightarrow[\mathbf{x}_{\text{init}} = \mathcal{F}(\mathbf{x}; \mathbf{\Omega})]{\text{KD-SemNOMA}} \mathbf{x}_{\text{init}} \xrightarrow[\mathbf{x}_{T'} \sim p(\mathbf{x}_{T'} \mid \mathbf{x}_{\text{init}})]{\text{error contraction}} \mathbf{x}_{T'} \xrightarrow[p_\theta(\mathbf{x}_{t-1} \mid \mathbf{x}_t), t = T' \to 1]{\text{reverse diffusion}} {\mathbf{x}_0}}_{\text{KDD-SemNOMA}}, \label{eq:refinement_pipeline}
	\vspace{-1.2mm}
\end{equation}
%\begin{equation}
%	\mathbf{x} \xrightarrow[\mathbf{x}_{\text{init}} = \mathcal{F}(\mathbf{x}; \mathbf{\Omega})]{\text{KD-SemNOMA}} \mathbf{x}_{\text{init}} \xrightarrow[\mathbf{x}_{T'} \sim p(\mathbf{x}_{T'} \mid \mathbf{x}_{\text{init}})]{\text{error contraction}} \mathbf{x}_{T'} \xrightarrow[p_\theta(\mathbf{x}_{t-1} \mid \mathbf{x}_t), t = T' \to 1]{\text{reverse diffusion}} {\mathbf{x}_0}, \label{eq:refinement_pipeline}
%\end{equation}
where $\mathcal{F}(\cdot; \mathbf{\Omega})$ represents the KD-SemNOMA model with trainable parameters $\mathbf{\Omega}$, $\mathbf{x}_{T'}$ is an intermediate diffused state, and ${\mathbf{x}_0}$ approximates $\mathbf{x}$.

\textbf{Obtaining $\mathbf{x}_{T'}$ via error contraction:} Defining the predicted error in KD-SemNOMA as $\mathbf{e} = \mathbf{x} - \mathbf{x}_{\text{init}}$. Using $\mathbf{x}_{\text{init}}$, we sample $\mathbf{x}_{T'}$ via the forward diffusion process
\begin{equation}
	\begin{aligned}
		\mathbf{x}_{T'} & =\sqrt{\alpha_{T'}} \mathbf{x}_{\text{init}}+\sqrt{\left(1-\alpha_{T'}\right)} \boldsymbol{\zeta} \\
		& =\sqrt{\alpha_{T'}}\mathbf{x} -\sqrt{\alpha_{T'}} \mathbf{e}+\sqrt{\left(1-\alpha_{T'}\right)} \boldsymbol{\zeta},
	\end{aligned}
	\label{eq:error_contraction}
	\vspace{-1.2mm}
\end{equation}
where $\boldsymbol{\zeta} \sim \mathcal{N}( 0, \mathbf{I}_m)$. $\alpha_{T'} = \prod_{t=1}^{T'} (1 - \beta_t)$, and $\beta_t$ denotes the predefined noise variance at timestep $t$. The predicted error $\mathbf{e}$ is contracted by a factor of $\sqrt{\alpha_{T'}}$ (where $0 < \sqrt{\alpha_{T'}} < 1$), reducing the impact of artifacts introduced during KD-SemNOMA reconstruction.

\textbf{Sampling with diffusion priors:} Starting from $\mathbf{x}_{T'}$, we employ a reverse diffusion process using a pre-trained transition kernel
\begin{equation}
	\mathbf{x}_{t-1} \sim p_\theta(\mathbf{x}_{t-1} \mid \mathbf{x}_t), \quad t = {T'}, {T'-1}, \dots, 1, \label{eq:reverse_sampling}
	\vspace{-1.5mm}
\end{equation}
where $p_\theta(\mathbf{x}_{t-1} \mid \mathbf{x}_t) = \mathcal{N}(\mathbf{x}_{t-1}; \boldsymbol{\mu}_\theta(\mathbf{x}_t, t), \sigma_t^2 \mathbf{I}_m)$, with $\boldsymbol{\mu}_\theta$ and $\sigma_t^2$ from the DDIM (Eq.~\ref{eq:ddim_posterior}). This iterative sampling refines $\mathbf{x}_{T'}$ into ${\mathbf{x}}_0$, leveraging the diffusion model's generative priors to enhance realism and detail of images.

To ensure fidelity across diverse multi-user images, we adopt the DDIM sampler with a randomness parameter $\eta = 0.5$. This approach mitigates excessive stochasticity, aligning $\mathbf{x}_{0}$ closely with $\mathbf{x}$ while preserving natural image characteristics. Additionally, we set the starting timestep $T' = 200$, enabling accelerated sampling compared to DDPM's typical $T \approx 1000$ steps, thus improving computational efficiency while maintaining high-quality refinement. In practice, we speed up the inference four times following \cite{nichol2021improved}, and thus sample 50 steps for each testing image. The complete inference procedure is detailed in Algorithm~\ref{alg:diffusion}.

The posterior $p(\mathbf{x}_0 | \mathbf{x}_{\text{init}})$ is implicitly constructed through the combination of the forward diffusion process $p(\mathbf{x}_{T'} | \mathbf{x}_{\text{init}})$ and the reverse sampling process $p_\theta(\mathbf{x}_{t-1} | \mathbf{x}_t)$. Such an issue is more prominent under stringent bandwidth constraints or low SNR conditions. 
\begin{algorithm}[!t]
	\caption{Inference procedure for diffusion model-based image refinement}
	\label{alg:diffusion}
	\begin{algorithmic}[1]
		\Require Test dataset $\mathbf{x}_{\text{test}} = \{ \mathbf{x}_{\text{init}}^{(1)}, \dots, \mathbf{x}_{\text{init}}^{(N)} \}$, noise schedule $\{ \beta_t \}_{t=1}^{T'}$, variance schedule $\sigma_t^2$
		\Ensure Enhanced multi-user images ${\mathbf{x}_0} = \{\mathbf{x}_0^{(1)}, \dots, \mathbf{x}_0^{(N)} \}$
		
		% ====== Inference Phase ======
		\Statex \textbf{Inference Phase (Enhancement with Pretrained Diffusion Model)}
		\State Input the initial reconstruction of KD-SemNOMA \( \mathbf{x}_{\text{init}} = \{ \mathbf{x}_{\text{init}}^{(1)}, \dots, \mathbf{x}_{\text{init}}^{(N)} \} \)
		%\State Reshape \(\mathbf{x}_{\text{init}} \in \mathbb{R}^{B \times N \times C \times H \times W}\) to \(\mathbf{x}_{\text{init}}^{\text{flat}} \in \mathbb{R}^{(B \cdot N) \times C \times H \times W}\)
		{\color{black}\State Reshape \(\mathbf{x}_{\text{init}} = \{ \mathbf{x}_{\text{init}}^{(i)} \}_{i=1}^{N}\) to \(\mathbf{x}_{\text{init}}^{\text{flat}} \in \mathbb{R}^{(B \cdot N) \times C \times H \times W}\)}
		\State Sample intermediate state via diffusion process according to Eq.~\ref{eq:error_contraction}:
		\For{\( t = T' \) down to \( 1 \)}
		%\State Predict noise \( \epsilon_\theta(x_t, t) \)
		\State Reverse sampling
		\Statex \hspace{1em} \( \mathbf{x}_{t-1} = \sqrt{\alpha_{t-1}} \mathbf{x}_{\text{init}} + \sqrt{1 - \alpha_{t-1}} \epsilon_\theta + \sigma_t \Xi \), \( \Xi \sim \mathcal{N}(0, \mathbf{I}_m) \)
		\EndFor
		%\State Reshape \({\mathbf{x}_{0}}^{\text{flat}} = \{ \mathbf{x}_0^{(i)} \}_{i=1}^{B \times N}\) to \({\mathbf{x}_0} \in \mathbb{R}^{B \times N \times C \times H \times W}\)
		\color{black}\State Reshape \({\mathbf{x}_{0}}^{\text{flat}} \in \mathbb{R}^{(B \cdot N) \times C \times H \times W}\) to \({\mathbf{x}_0} = \{ \mathbf{x}_{0}^{(i)} \}_{i=1}^{N}\)
		\color{black}\State \Return \({\mathbf{x}_0} = \{ \mathbf{x}_0^{(1)}, \dots, \mathbf{x}_0^{(N)} \} \)
	\end{algorithmic}
\end{algorithm}
%\vspace{-3mm}
\section{Training and Evaluation} \label{Simulation}
\subsection{Simulation Setup}
\subsubsection{Datasets}
For low-resolution datasets, we employ CIFAR-10 dataset \cite{krizhevsky2009learning} for training and testing. It consists of 50,000 training images and 10,000 testing images, each with the shape of $C \times W \times H = 3 \times 32 \times 32$. For high-resolution datasets, we employ FFHQ dataset \cite{karras2019style} which comprises $70,000$ high-quality face images with a native resolution of $C \times W \times H = 3 \times 1024 \times 1024$, containing significant variations in age, race, and image background. For training, we down-sample the FFHQ images to $C \times W \times H = 3 \times 256 \times 256$ using bi-cubic down-sampling.
In order to build dataset for multi-user scenarios, we randomly sample $200,000$ samples for the training set and dynamically group the samples from the original dataset into user specific batches. For the testing set, we divide the datasets into $N$ non-overlapping groups and concatenate them into the multi-user dataset. Simultaneously, $\gamma$ is uniformly sampled from $[0, 20]$ dB, and the Rayleigh parameters $(a_i, \phi_i)$ are randomly generated to simulate diverse channel conditions. 

\subsubsection{Benchmark Schemes}
For the performance evaluation, we compare our proposed scheme with the following baseline schemes.
\begin{itemize}
	{\color{black}\item \textbf{BPG+LDPC+QAM+SIC}:
		In this case, the conventional scheme performs the source coding and channel coding separately. Better Portable Graphics (BPG) and low-density parity-check (LDPC) are employed for the source and channel coding. The $1/2$ code rate LDPC code is employed with information block length of 4096 bits and codeword length of 8192 bits. The modulation scheme is 4 quadrature amplitude modulation (QAM). For multi-user detection at the receiver, the successive interference cancellation (SIC) algorithm \cite{ding2020unveiling} is employed.}
	%\item \textbf{JPEG+LDPC+QAM+SIC}:
	%In this case, the conventional scheme performs the source coding and channel coding separately. Joint photographic experts group (JPEG) and low-density parity-check (LDPC) are employed for the source and channel coding. The LDPC code employs $3/4$ code rate. The modulation scheme is 4 quadrature amplitude modulation (QAM). For multi user detection at the receiver, the successive interference cancellation (SIC) algorithm \cite{ding2020unveiling} is employed. 
	\item \textbf{DeepJSCC-NOMA}: 
	In this case, we adopt the attention based DeepJSCC (ADJSCC) scheme \cite{yilmaz2023distributed} for multi-user NOMA wireless images transmission, which is trained on CIFAR-10 and FFHQ-256. 
	\item \textbf{SemOMA}: 
	In this case, we assume that multi-user semantic features are transmitted through orthogonal multiple access (OMA) channels with equal power. For this benchmark, we consider two configurations of $\rho$: one where SemOMA incurs the same transmission overhead as SemNOMA, and another where it incurs twice the transmission overhead.
	\item \textbf{SemNOMA-CGAN}:
	In this case, borrow from GAN decomposition method \cite{zou2020deep}, we combine the proposed SemNOMA with conditional GAN (CGAN) network to enhance the perceptual quality of the multi-user restored images. We utilize the U-Net \cite{ronneberger2015unet} based generator and use the combination of MAE loss, learned perceptual image patch similarity (LPIPS) loss, and GAN loss to train the CGAN network. {\color{black}Detailed training implementation of SemNOMA-CGAN is provided in Section \ref{training_details}.}
	{\color{black}\item \textbf{SemNOMA-DiffBIR}:
	Build upon the two-stage framework of DiffBIR \cite{lindiffbir}, we integrate the SemNOMA with the pretrained diffusion-based generation module from DiffBIR, which introduces the IRControlNet that leverages Stable Diffusion \cite{rombachhigh} prior for realistic image restoration. For inference, LLaVA-7B \cite{liuvisual} is applied to the SemNOMA initial output to generate a descriptive caption, which is then concatenated with a fixed high-quality positive prompt. The resulting text prompt, together with the initial output image as condition, is fed into the Stable Diffusion model equipped with IRControlNet for generative sampling.} 
\end{itemize}

\subsubsection{Training and Inference Details} \label{training_details}

We train the KD-SemNOMA framework following Algorithm~\ref{alg:kd}. Specifically, we employ the AdamW optimizer \cite{loshchilov2017decoupled} with a learning rate of $1\times10^{-4}$ and a weight decay of $1\times10^{-4}$. Both the teacher and student models are trained using the batch size of 32. To mitigate overfitting, an early stopping mechanism is implemented, terminating training when no improvement in validation performance is observed over 10 consecutive epochs. During student model training, the loss weights are set to $\lambda_1 = 10$, $\lambda_2 = 100$, and $\lambda_3 = 1$ to balance the contributions of the restoration loss and distillation terms, as defined in Section~\ref{KD}. During inference, KD-SemNOMA relies solely on the student model, incurring no additional computational overhead compared to the proposed SemNOMA model as detailed in Section \ref{SemNOMA_arch}. For DDIM inference, the first-stage output $\mathbf{x}_{\text{init}} = \hat{\mathbf{x}}$ is refined through iterative denoising (Eq.~\ref{eq:refinement_pipeline}), using $\eta = 0.5$ and $T' = 200$ steps to maximize image perceptual performance. {\color{black}For the SemNOMA-CGAN baseline, followed \cite{zou2020deep}, we employ a CGAN with a U-Net based generator and a PatchGAN discriminator. The CGAN is trained using the outputs from the pre-trained KD-SemNOMA model as conditional input. During training, the KD-SemNOMA weights remain frozen while the CGAN components are optimized through AdamW optimizer over 100 epochs with a composite loss function where the weights are set to 1 for MAE, 0.1 for LPIPS, and $1\times10^{-4}$ for the adversarial loss.}

For datasets of different resolutions, the encoder and decoder network parameters of the proposed ConvNeXt-based DeepJSCC are detailed in Table~\ref{tab:encoder_decoder_cifar10} for CIFAR-10 and Table~\ref{tab:encoder_decoder_ffhq256} for FFHQ-256. The proposed framework and all comparative methods are implemented in PyTorch and executed on an NVIDIA 4090 GPU server.
\begin{table}[ht]
	\centering
	\caption{Parameters setting of DeepJSCC encoder and decoder for the image resolution of 32×32}
	\label{tab:encoder_decoder_cifar10}
	
	% ====== Encoder Parameters ======
	\begin{tabular}{p{0.18\textwidth}|p{0.262\textwidth}}
		\hline
		\textbf{Encoder Layer} & \textbf{Parameters} \\
		\hline
		Input & Input shape $3 \times 32 \times 32$  \\
		Stem (downsample) & Conv2D ($C \to 96$, kernel=4, stride=2) \\
		Downsample Layer 1 & Conv2D ($96 \to 192$, kernel=2, stride=2) \\
		Stage 1 (ConvNeXt Blocks) & Depth=3, Dim=96, Drop Path Rate=0.1 \\
		Stage 2 (ConvNeXt Blocks) & Depth=3, Dim=192, Drop Path Rate=0.1 \\
		Output Layer & Conv2D ($192 \to 16$, kernel=1) \\
		\hline
	\end{tabular}
	
	\vspace{0.5em} % 添加间距
	
	% ====== Decoder Parameters ======
	\begin{tabular}{p{0.18\textwidth}|p{0.262\textwidth}}
		\hline
		\textbf{Decoder Layer} & \textbf{Parameters} \\
		\hline
		Input & Input shape $16 \times 8 \times 8$ \\
		Input Layer & Conv2D ($16 \to 192$, kernel=1) \\
		Stage 1 (ConvNeXt Blocks) & Depth=3, Dim=192, Drop Path Rate=0.1 \\
		Upsample Layer 1 & Upsample ($8 \times 8 \to 16 \times 16$, factor=2) \\
		Stage 2 (ConvNeXt Blocks) & Depth=3, Dim=96, Drop Path Rate=0.1 \\
		Upsample Layer 2 & Upsample ($16 \times 16 \to 32 \times 32$, factor=2) \\
		Output Layer & Conv2D ($96 \to 3$, kernel=1) \\
		\hline
	\end{tabular}
\end{table}
\vspace{-1em}
\begin{table}[ht]
	\centering
	\caption{Parameters setting of DeepJSCC encoder and decoder for the image resolution of 256×256}
	\label{tab:encoder_decoder_ffhq256}
	
	% ====== Encoder Parameters ======
	\begin{tabular}{p{0.18\textwidth}|p{0.262\textwidth}}
		\hline
		\textbf{Encoder Layer} & \textbf{Parameters} \\
		\hline
		Input & Input shape $3 \times 256 \times 256$  \\
		Stem (downsample) & Conv2D ($3 \to 96$, kernel=4, stride=2) \\
		Downsample Layer 1 & Conv2D ($96 \to 192$, kernel=2, stride=2) \\
		Stage 1 (ConvNeXt Blocks) & Depth=2, Dim=96, Drop Path Rate=0.1 \\
		Downsample Layer 2 & Conv2D  ($192 \to 384$, kernel=2, stride=2) \\
		Stage 2 (ConvNeXt Blocks) & Depth=2, Dim=192, Drop Path Rate=0.1 \\
		Downsample Layer 3 & Conv2D ($384 \to 768$, kernel=2, stride=2) \\
		Stage 3 (ConvNeXt Blocks) & Depth=6, Dim=384, Drop Path Rate=0.1 \\
		Stage 4 (ConvNeXt Blocks) & Depth=2, Dim=768, Drop Path Rate=0.1 \\
		Output Layer & Conv2D ($768 \to 32$, kernel=1) \\
		\hline
	\end{tabular}
	
	\vspace{0.5em} % 添加间距
	
	% ====== Decoder Parameters ======
	\begin{tabular}{p{0.18\textwidth}|p{0.262\textwidth}}
		\hline
		\textbf{Decoder Layer} & \textbf{Parameters} \\
		\hline
		Input & Input shape $32 \times 16 \times 16$ \\
		Input Layer & Conv2D ($32 \to 768$, kernel=1) \\
		Stage 1 (ConvNeXt Blocks) & Depth=2, Dim=768, Drop Path Rate=0.1 \\
		Upsample Layer 1 & Upsample ($16 \times 16 \to 32 \times 32$) \\
		Stage 2 (ConvNeXt Blocks) & Depth=6, Dim=384, Drop Path Rate=0.1 \\
		Upsample Layer 2 & Upsample ($32 \times 32 \to 64 \times 64$) \\
		Stage 3 (ConvNeXt Blocks) & Depth=2, Dim=192, Drop Path Rate=0.1 \\
		Upsample Layer 3 & Upsample ($64 \times 64 \to 128 \times 128$) \\
		Stage 4 (ConvNeXt Blocks) & Depth=2, Dim=96, Drop Path Rate=0.1 \\
		Upsample Layer 4 & Upsample ($128 \times 128 \to 256 \times 256$) \\
		Output Layer & Conv2D ($96 \to 3$, kernel=1) \\
		\hline
	\end{tabular}
	\vspace{-3mm}
\end{table}
\subsubsection{Performance Metrics}
We adopt standard distortion metrics such as PSNR and structural similarity index (SSIM) to measure image fidelity concerning pixel intensity and structural details, respectively. To evaluate the perceptual quality of reconstructed images, we employ LPIPS \cite{zhang2018unreasonable}, computed with VGG16 \cite{simonyan2014very} pretrained on ImageNet and Fréchet inception distance (FID) scores \cite{heusel2017gans}, using Inception V3 to measure feature similarity. FID assesses visual quality by calculating the statistical similarity between the original image set and the reconstructed image set.
%\vspace{-2mm}
\subsection{Performance of Knowledge Distillation-Based SemNOMA for Image Transmission} \label{experiment of KD-SemNOMA}
\begin{figure*}[!t]
	\centering
	\begin{subfigure}[t]{0.5\columnwidth}
		\centering
		\includegraphics[width=\linewidth]{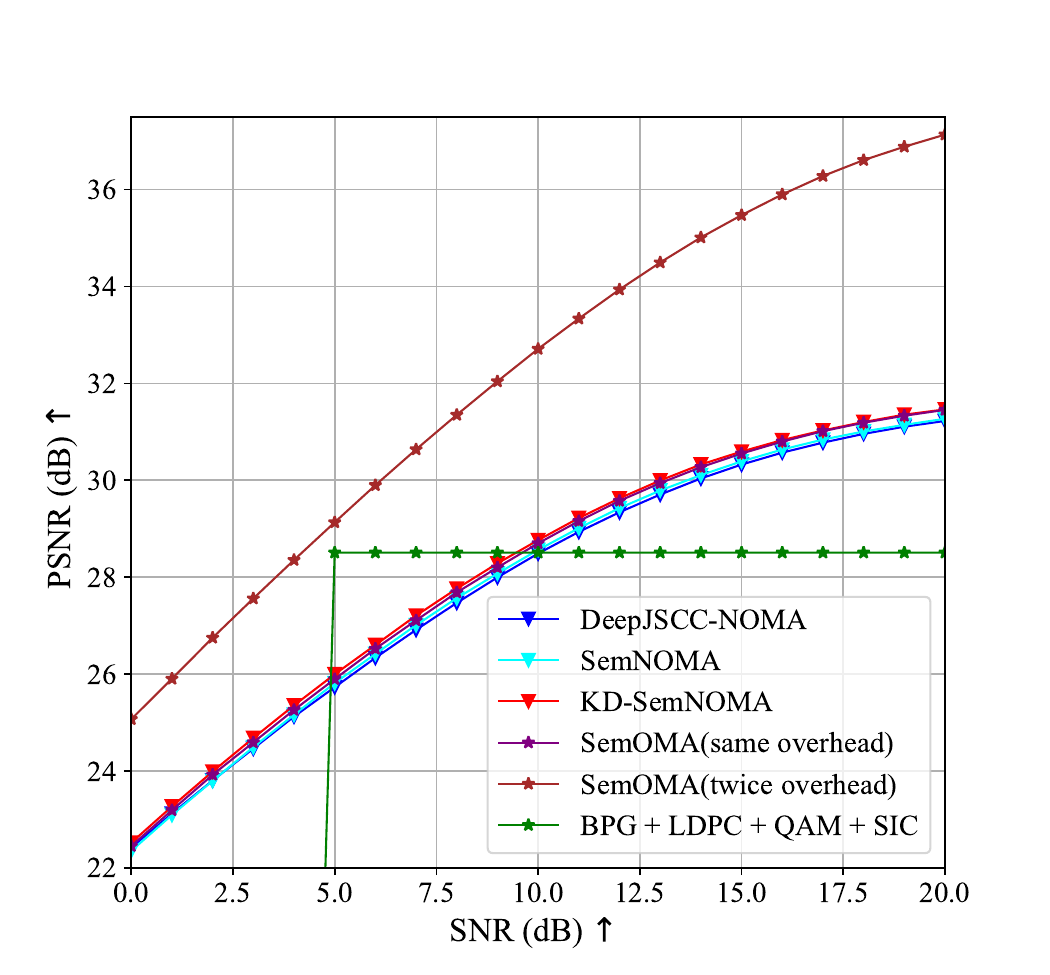}
		\captionsetup{font={footnotesize}}
		\caption{PSNR vs. SNR ($\rho=1/6$)}
		\label{fig1a}
	\end{subfigure}
	\hfill
	\begin{subfigure}[t]{0.5\columnwidth}
		\centering
		\includegraphics[width=\linewidth]{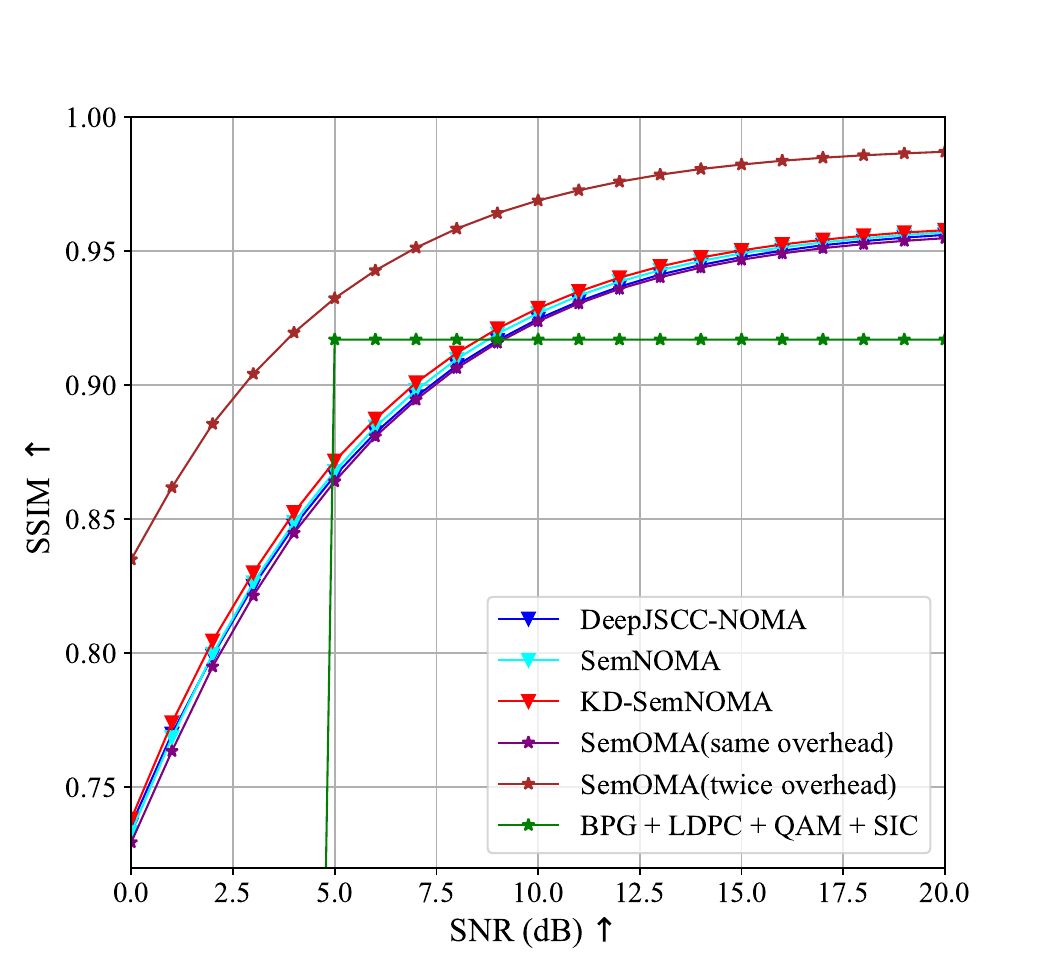}
		\captionsetup{font={footnotesize}}
		\caption{SSIM vs. SNR ($\rho=1/6$)}
		\label{fig1b}
	\end{subfigure}
	\hfill
	\begin{subfigure}[t]{0.5\columnwidth}
		\centering
		\includegraphics[width=\linewidth]{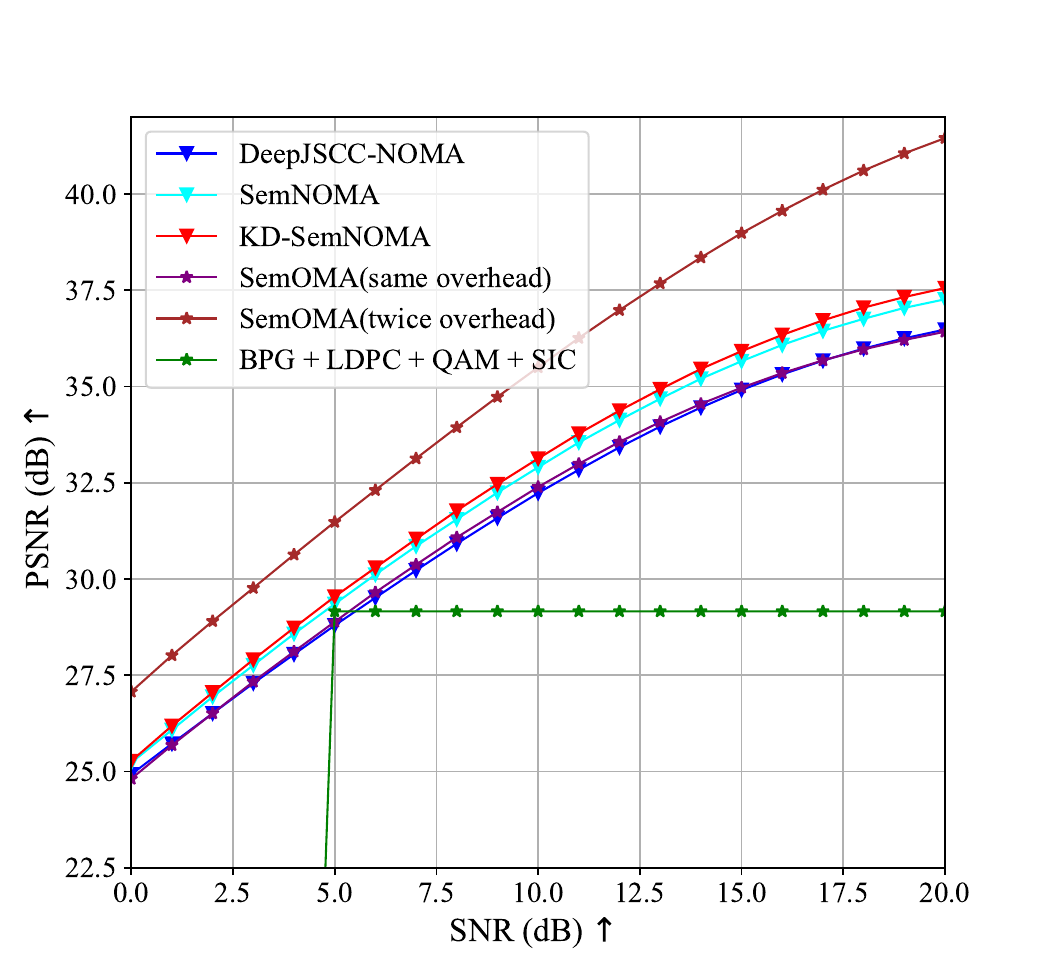}
		\captionsetup{font={footnotesize}}
		\caption{PSNR vs. SNR ($\rho=1/3$)}
		\label{fig4a}
	\end{subfigure}
	\hfill
	\begin{subfigure}[t]{0.5\columnwidth}
		\centering
		\includegraphics[width=\linewidth]{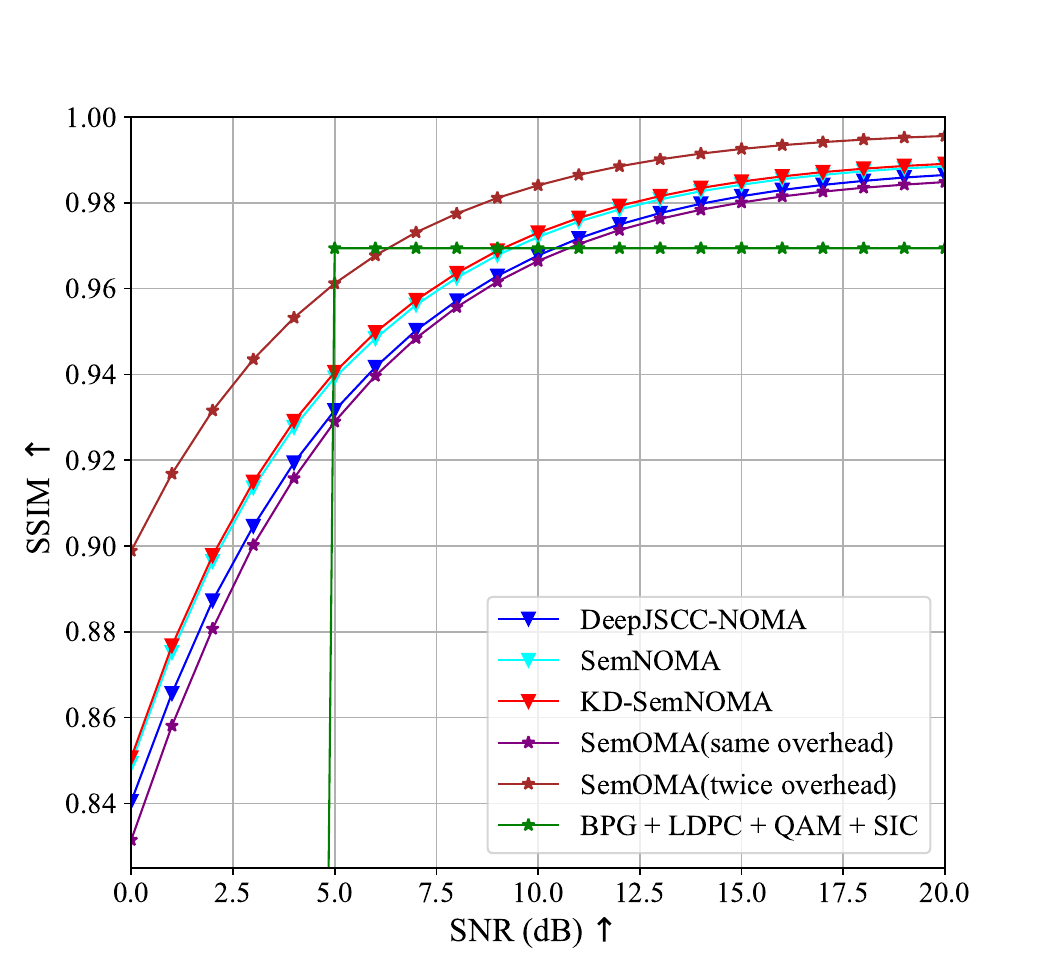}
		\captionsetup{font={footnotesize}}
		\caption{SSIM vs. SNR ($\rho=1/3$)}
		\label{fig4b}
	\end{subfigure}
	\captionsetup{font={footnotesize,color=black}}
	\caption{Performance comparison on CIFAR-10 dataset under AWGN channel (2UE, $\rho=1/6$ and  $\rho=1/3$). Subfigures (a) and (b) show PSNR and SSIM versus SNR for $\rho=1/6$, while subfigures (c) and (d) show PSNR and SSIM for $\rho=1/3$.}
	\label{fig:performance_awgn_cifar10}
	\vspace{-4mm}
\end{figure*}
We evaluate the proposed \textbf{KD-SemNOMA} framework for image transmission, comparing it against baselines (i.e., {\color{black}\textbf{BPG+LDPC+QAM+SIC}}, \textbf{DeepJSCC-NOMA}) and semantic orthogonal transmission schemes (i.e., \textbf{SemOMA}) on CIFAR-10 and FFHQ-256 datasets under AWGN and Rayleigh fading channels, with 2 UE and compression ratios $\rho$ of 1/6 and 1/3 for CIFAR-10, and 1/48 for FFHQ-256. The ConvNeXt-based \textbf{KD-SemNOMA}, optimized via Algorithm~\ref{alg:kd}, leverages knowledge distillation stragety to enhance pixel-level fidelity (i.e., PSNR and SSIM), as detailed below.

We first conduct simulation comparison between the proposed \textbf{KD-SemNOMA} and the baseline scheme. As shown in Fig.~\ref{fig:performance_awgn_cifar10}, we evaluate the image reconstruction performance in terms of PSNR and SSIM on the CIFAR-10 dataset under the AWGN channel with a compression ratio $\rho$ of $1/6$ and $1/3$ across various $\gamma$ conditions. The results indicate that the {\color{black}\textbf{BPG+LDPC+QAM+SIC}} method suffers from a cliff effect at low SNR, leading to a significant degradation in image reconstruction. In our simulations, we employs non-equal power allocation ({\color{black}power factors of 0.8 and 0.2 for the two users}) in {\color{black}\textbf{BPG+LDPC+QAM+SIC}} to enable feasible SIC decoding, yet it still fails to overcome the interference-limited performance floor. Additionally, the conventional NOMA scheme with SIC detection algorithm fails to effectively separate signals and mitigate interference in equal power allocation scenarios, resulting in poor PSNR and SSIM performance. In contrast, the proposed deep learning-based multi-user image transmission framework performs robustly at low SNR and demonstrates strong adaptability to arbitrary power allocation scenarios. Compared to the baseline \textbf{DeepJSCC-NOMA} framework based on the ResNet architecture, our proposed \textbf{SemNOMA} scheme, leveraging the ConvNeXt network, outperforms the baseline across various $\gamma$ conditions, validating the effectiveness of the ConvNeXt network in extracting deep image features. 

Furthermore, through optimization with knowledge distillation, \textbf{KD-SemNOMA} achieves an approximate PSNR gain of $0.2 \sim 0.3$dB in image reconstruction, with further improvements in SSIM. This demonstrating that knowledge distillation effectively transfers interference-free feature knowledge from the teacher model to the student model. As a result, the student network’s single-user decoding branch can perform independent decoding with near-interference-free features, thus enhancing the overall image reconstruction performance. We also compared our scheme with \textbf{SemOMA} with the same overhead and twice the overhead. Compared with the \textbf{SemNOMA} scheme with the same overhead, the \textbf{SemNOMA} scheme can achieve PSNR performance gain of $0.2 \sim 0.8$dB, and compared with the \textbf{SemOMA} with twice the overhead, it also has certain performance potential, proving that \textbf{SemNOMA} can improve image reconstruction performance while reducing transmission overhead.
\begin{figure*}[!t]
	\centering
	\begin{subfigure}[t]{0.5\columnwidth}
		\centering
		\includegraphics[width=\linewidth]{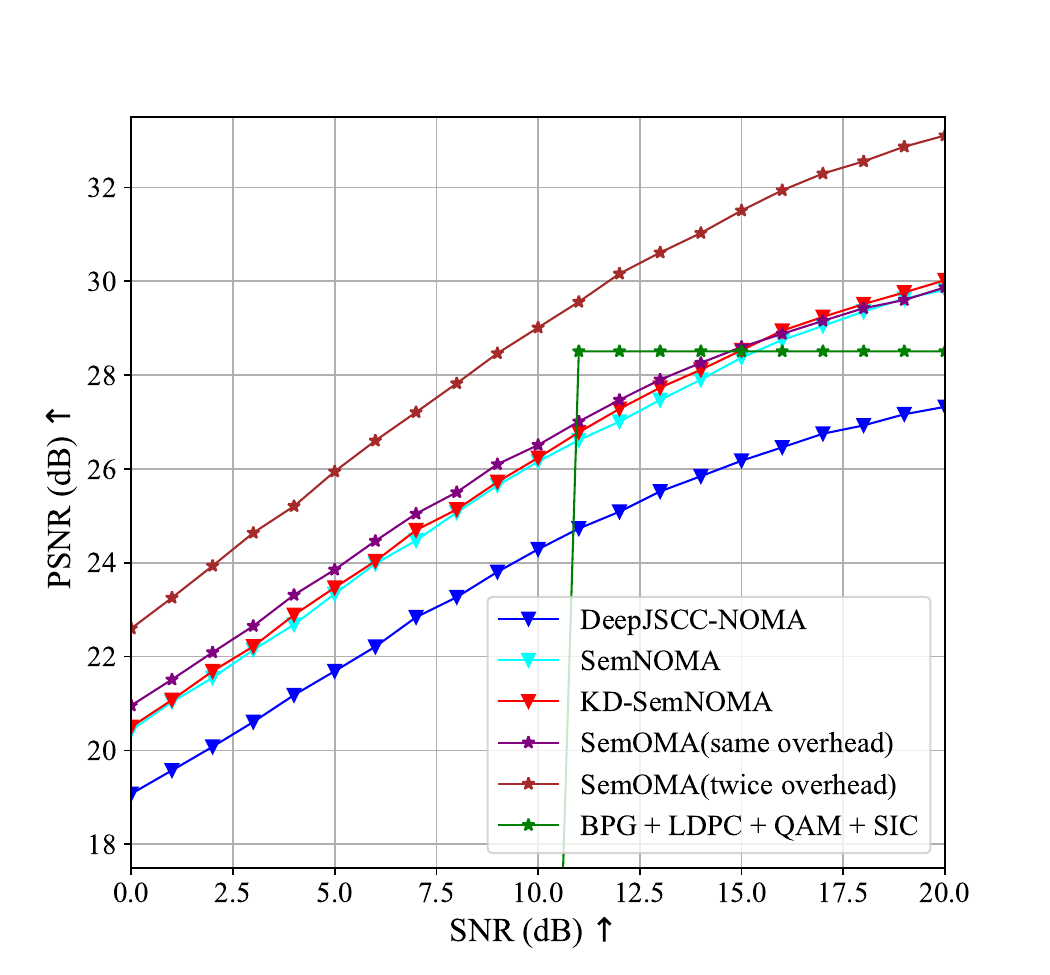}
		\captionsetup{font={footnotesize}}
		\caption{PSNR vs. SNR ($\rho=1/6$)}
		\label{fig2a}
	\end{subfigure}
	\hfill
	\begin{subfigure}[t]{0.5\columnwidth}
		\centering
		\includegraphics[width=\linewidth]{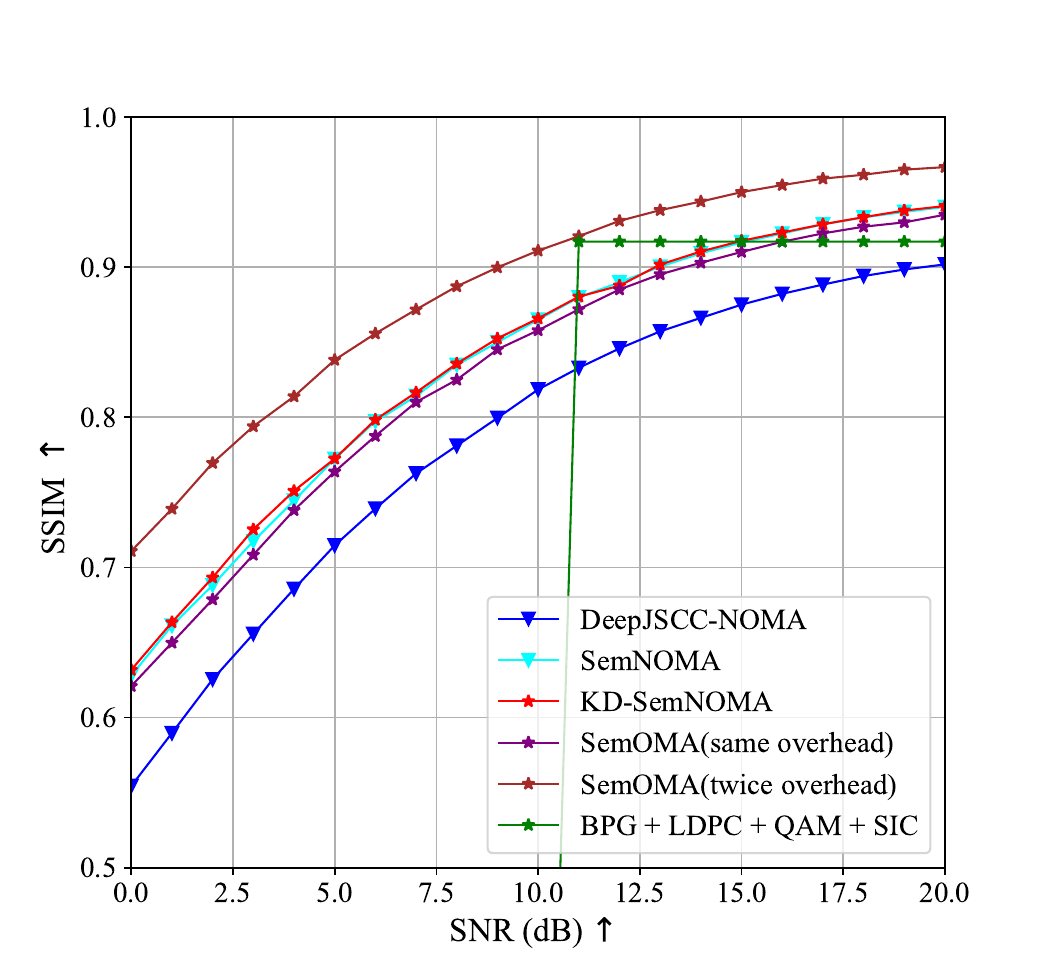}
		\captionsetup{font={footnotesize}}
		\caption{SSIM vs. SNR ($\rho=1/6$)}
		\label{fig2b}
	\end{subfigure}
	\hfill
	\begin{subfigure}[t]{0.5\columnwidth}
		\centering
		\includegraphics[width=\linewidth]{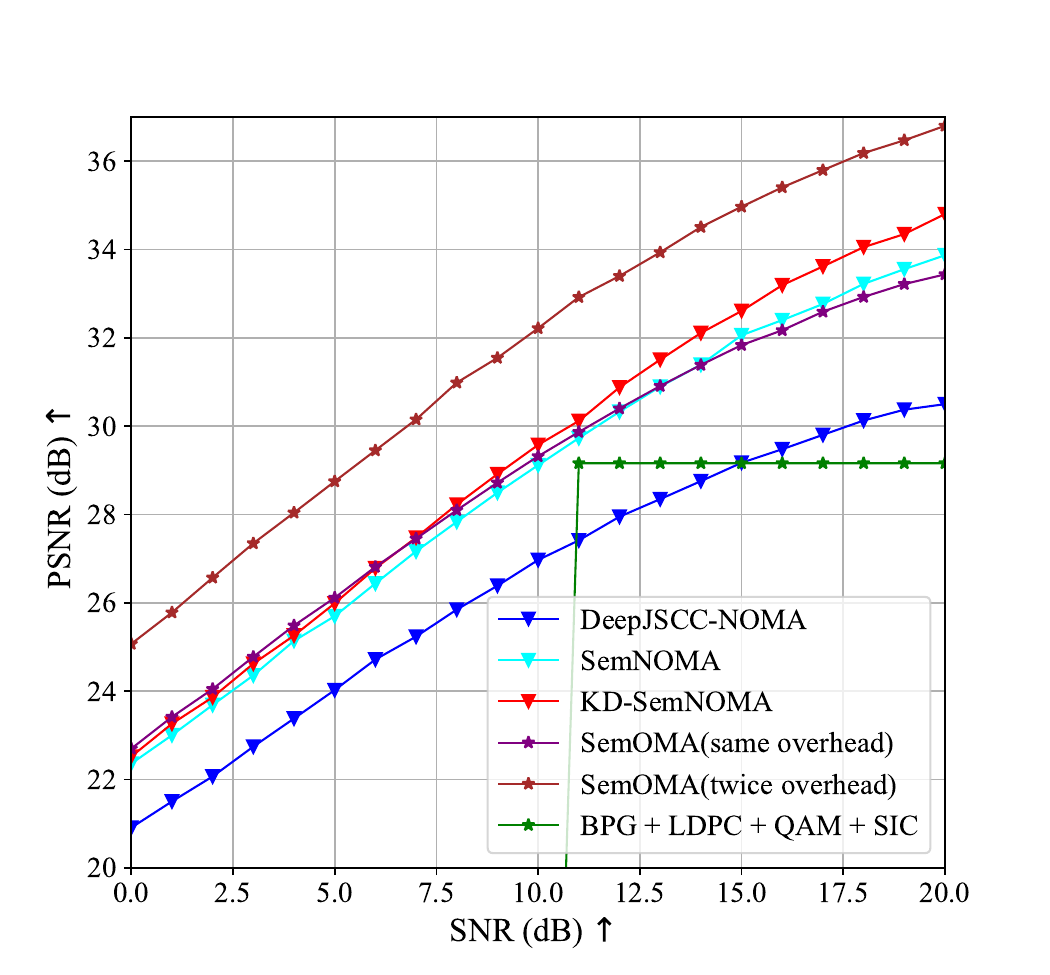}
		\captionsetup{font={footnotesize}}
		\caption{PSNR vs. SNR ($\rho=1/3$)}
		\label{fig3a}
	\end{subfigure}
	\hfill
	\begin{subfigure}[t]{0.5\columnwidth}
		\centering
		\includegraphics[width=\linewidth]{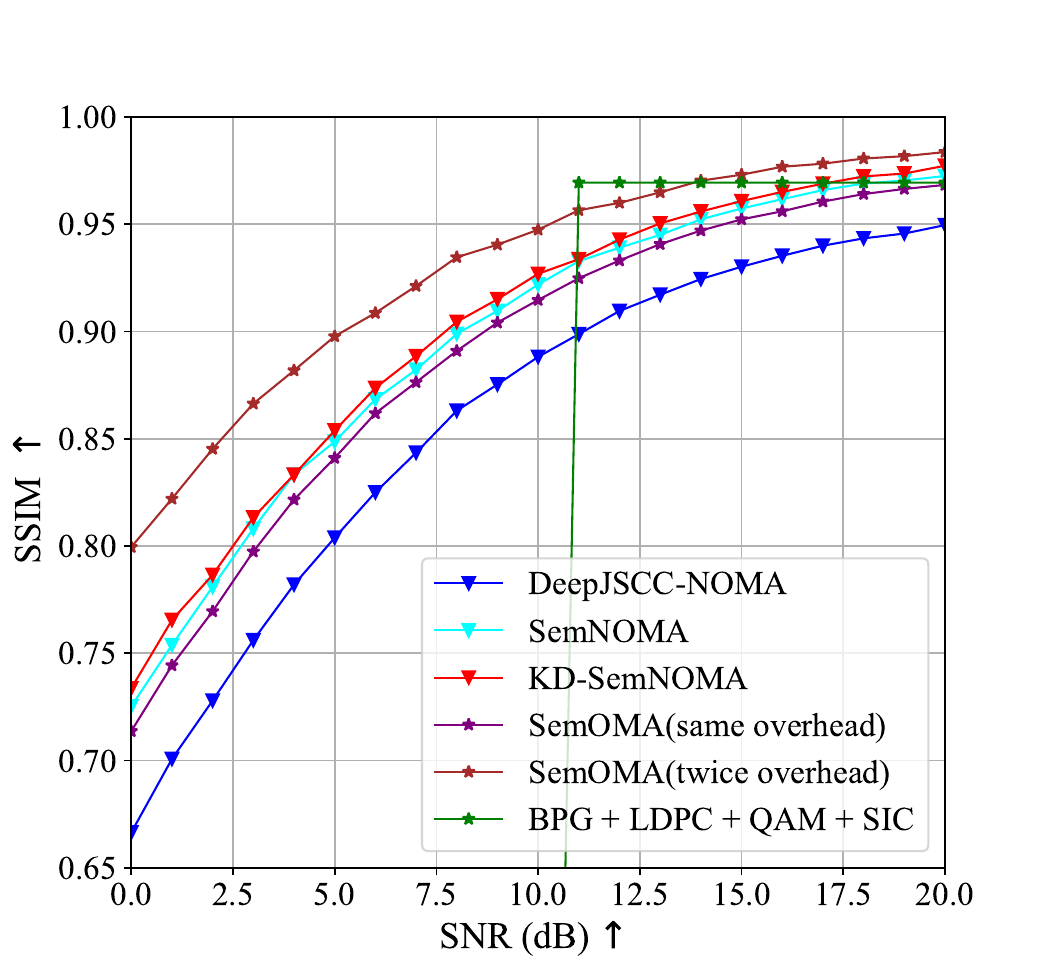}
		\captionsetup{font={footnotesize}}
		\caption{SSIM vs. SNR ($\rho=1/3$)}
		\label{fig3b}
	\end{subfigure}
	\captionsetup{font={footnotesize,color=black}}
	\caption{Performance comparison on CIFAR-10 dataset under Rayleigh fading channel (2UE,  $\rho=1/6$ and  $\rho=1/3$). Subfigures (a) and (b) show PSNR and SSIM versus SNR for $\rho=1/6$, while subfigures (c) and (d) show PSNR and SSIM for $\rho=1/3$.}
	\label{fig:performance_rayleigh_cifar10}
	\vspace{-4mm}
\end{figure*}

In Fig.~\ref{fig:performance_rayleigh_cifar10}, we simulate the image reconstruction PSNR and SSIM performance of different schemes under the Rayleigh channel with the compression ratios $\rho$ of $1/6$ and $1/3$ on the CIFAR-10 dataset across various $\gamma$ conditions. The results show that the enhanced AF-Module combines the SNR $\gamma$ and the amplitude $a$ and phase $\phi$ coefficients of the Rayleigh channel as channel side information input to the encoder and decoder of the model, enabling the network to learn channel characteristics and adapt to channel changes, thereby improving image reconstruction performance. Compared to the baseline \textbf{DeepJSCC-NOMA}, which only adapts to $\gamma$, the proposed framework with the enhanced AF-Module achieves a PSNR improvement of $1 \sim 2.5$dB when $\rho=1/6$ and $2 \sim 3$dB when $\rho=1/3$ , with SSIM also surpassing the baseline, demonstrating the robustness of the proposed framework across different channel conditions, such as AWGN and Rayleigh channel. Additionally, \textbf{KD-SemNOMA} under the Rayleigh channel provides a further gain of approximately $0.2$dB when $\rho=1/6$ and $0.2 \sim 0.8$dB when $\rho=1/3$, respectively. With equivalent transmission overhead, compared to \textbf{SemOMA}, \textbf{KD-SemNOMA} performs comparably to the orthogonal scheme at low SNR range in terms of PSNR, while outperforming it at high SNR range. For the SSIM metric, \textbf{KD-SemNOMA} consistently outperforms \textbf{SemOMA}, confirming its ability to deliver powerful image reconstruction performance improvement while conserving resource overhead in complex channel environments.
\begin{figure*}[!t]
	\centering
	\begin{subfigure}[t]{0.5\columnwidth}
		\centering
		\includegraphics[width=\linewidth]{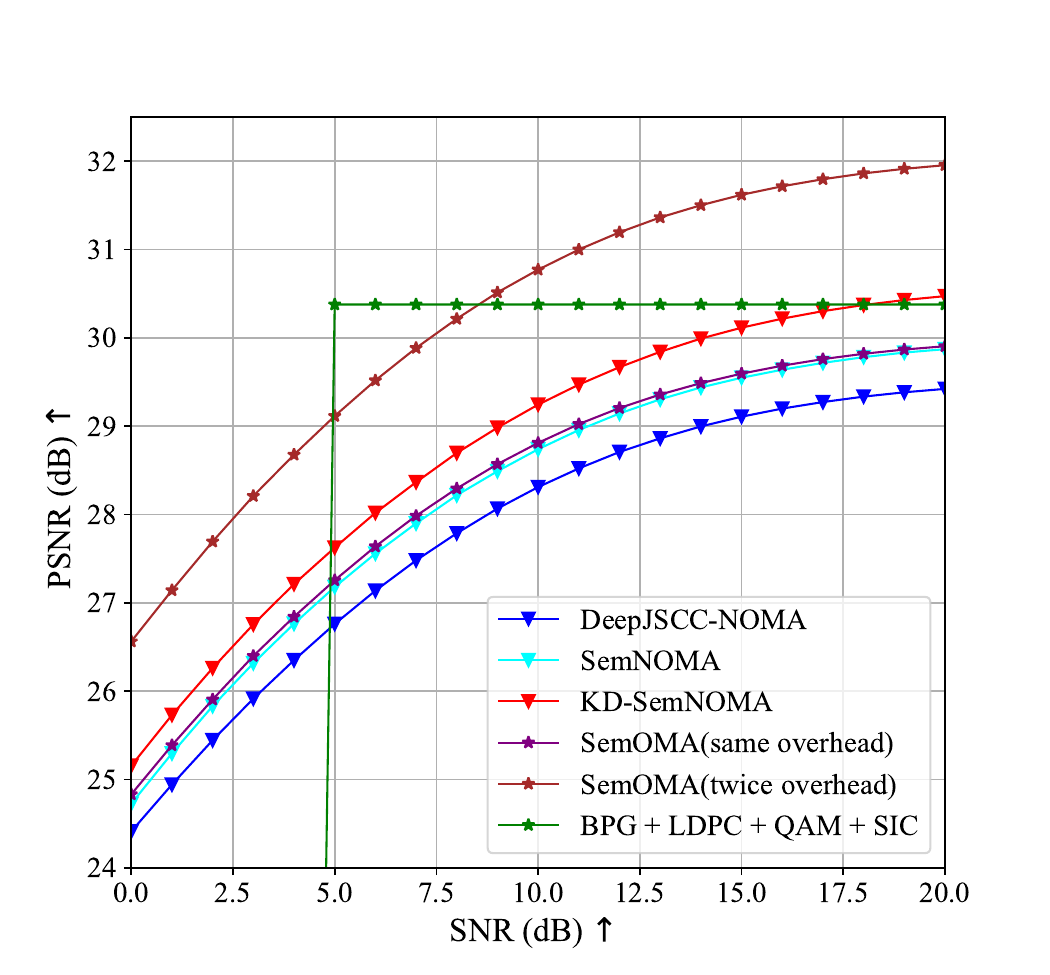}
		\captionsetup{font={footnotesize}}
		\caption{PSNR vs. SNR (AWGN)}
		\label{fig6a}
	\end{subfigure}
	\hfill
	\begin{subfigure}[t]{0.5\columnwidth}
		\centering
		\includegraphics[width=\linewidth]{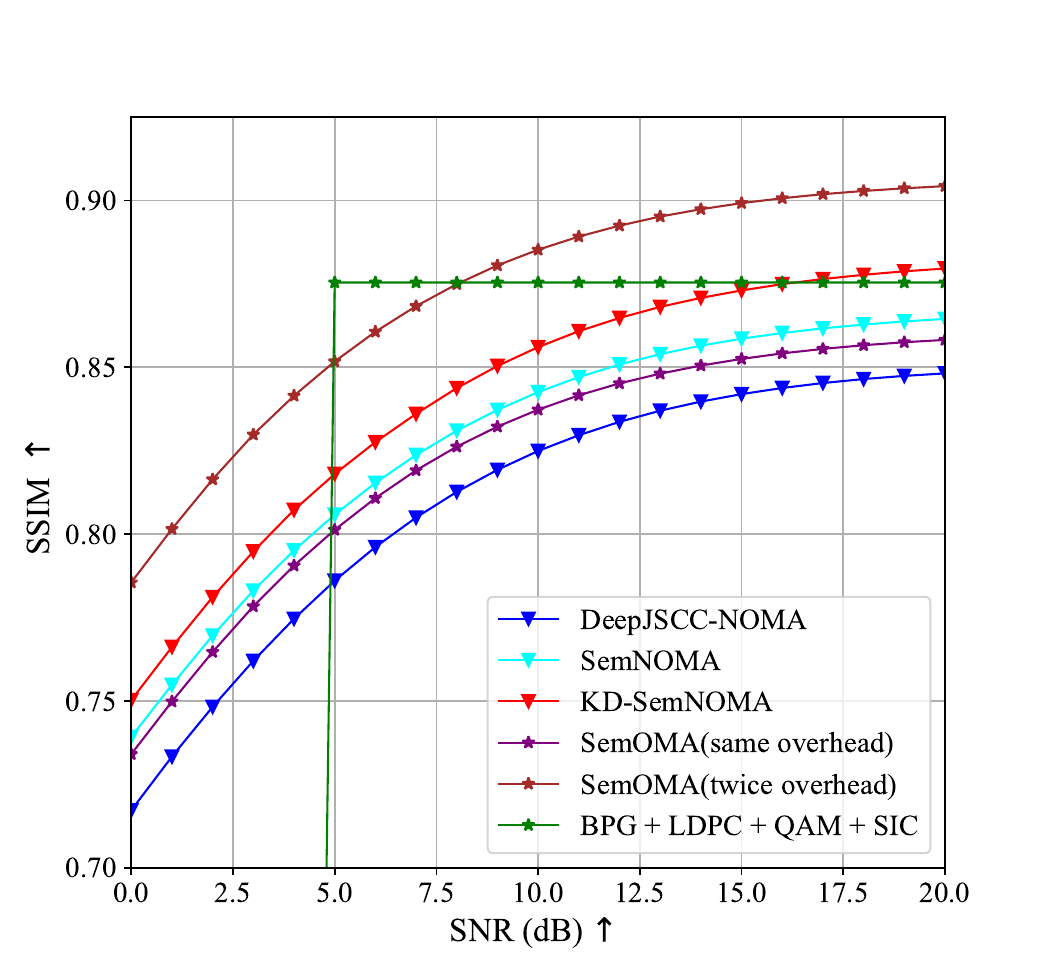}
		\captionsetup{font={footnotesize}}
		\caption{SSIM vs. SNR (AWGN)}
		\label{fig6b}
	\end{subfigure}
	\hfill
	\begin{subfigure}[t]{0.5\columnwidth}
		\centering
		\includegraphics[width=\linewidth]{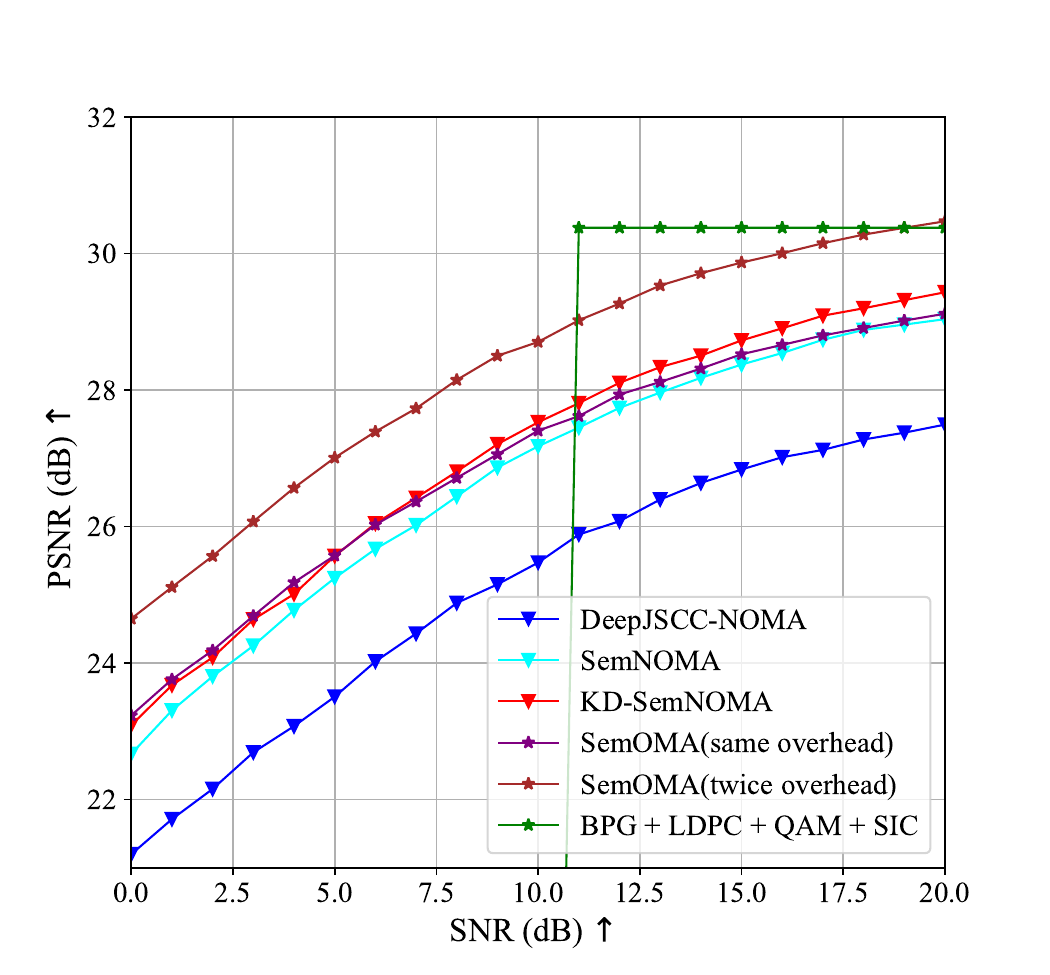}
		\captionsetup{font={footnotesize}}
		\caption{PSNR vs. SNR (Rayleigh)}
		\label{fig5a}
	\end{subfigure}
	\hfill
	\begin{subfigure}[t]{0.5\columnwidth}
		\centering
		\includegraphics[width=\linewidth]{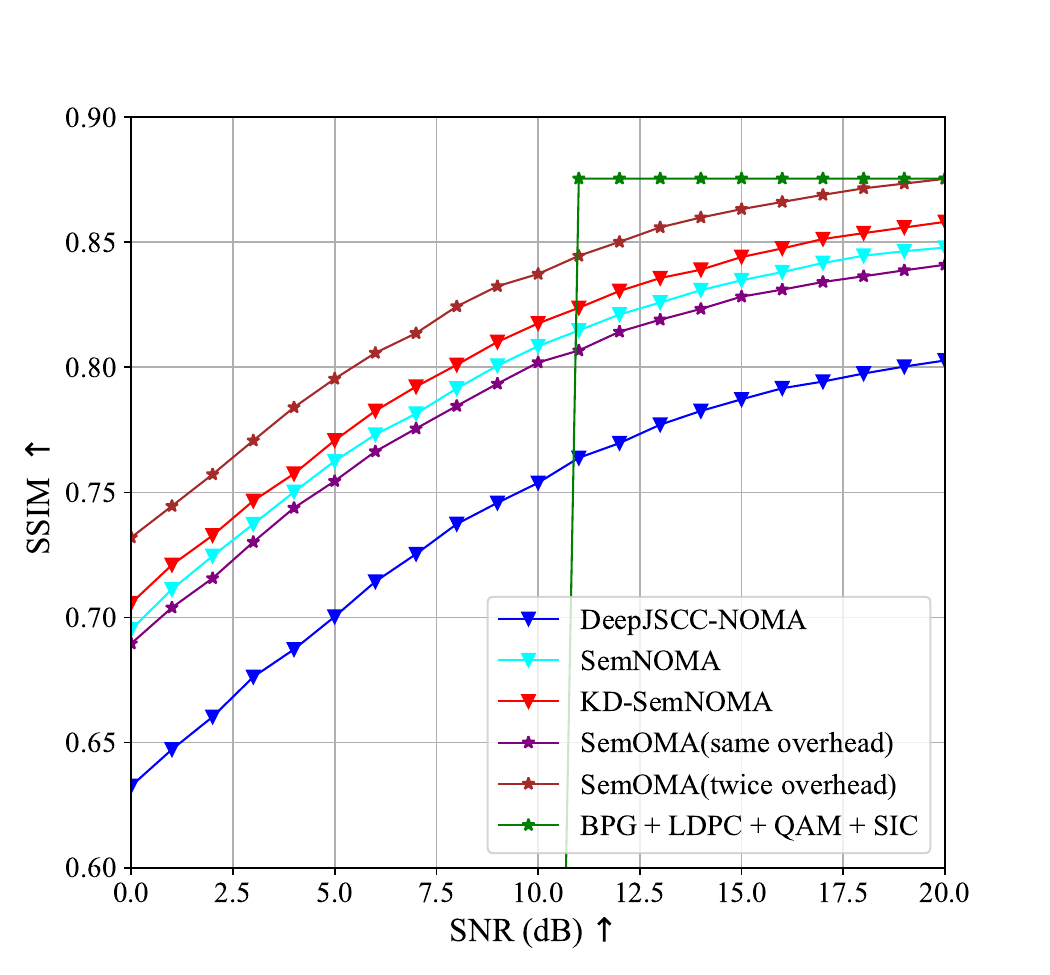}
		\captionsetup{font={footnotesize}}
		\caption{SSIM vs. SNR (Rayleigh)}
		\label{fig5b}
	\end{subfigure}
	\captionsetup{font={footnotesize,{color=black}}}
	\caption{Performance comparison on FFHQ-256 dataset under AWGN and Rayleigh fading channels (2UE, $\rho=1/48$). Subfigures (a) and (b) show PSNR and SSIM versus SNR under AWGN channel, while subfigures (c) and (d) show PSNR and SSIM under Rayleigh fading channel.}
	\label{fig:performance_ffhq_M32}
	\vspace{-4mm}
\end{figure*}
\begin{figure*}[!t]
	\centering
	\begin{subfigure}[t]{0.5\columnwidth}
		\centering
		\includegraphics[width=\linewidth]{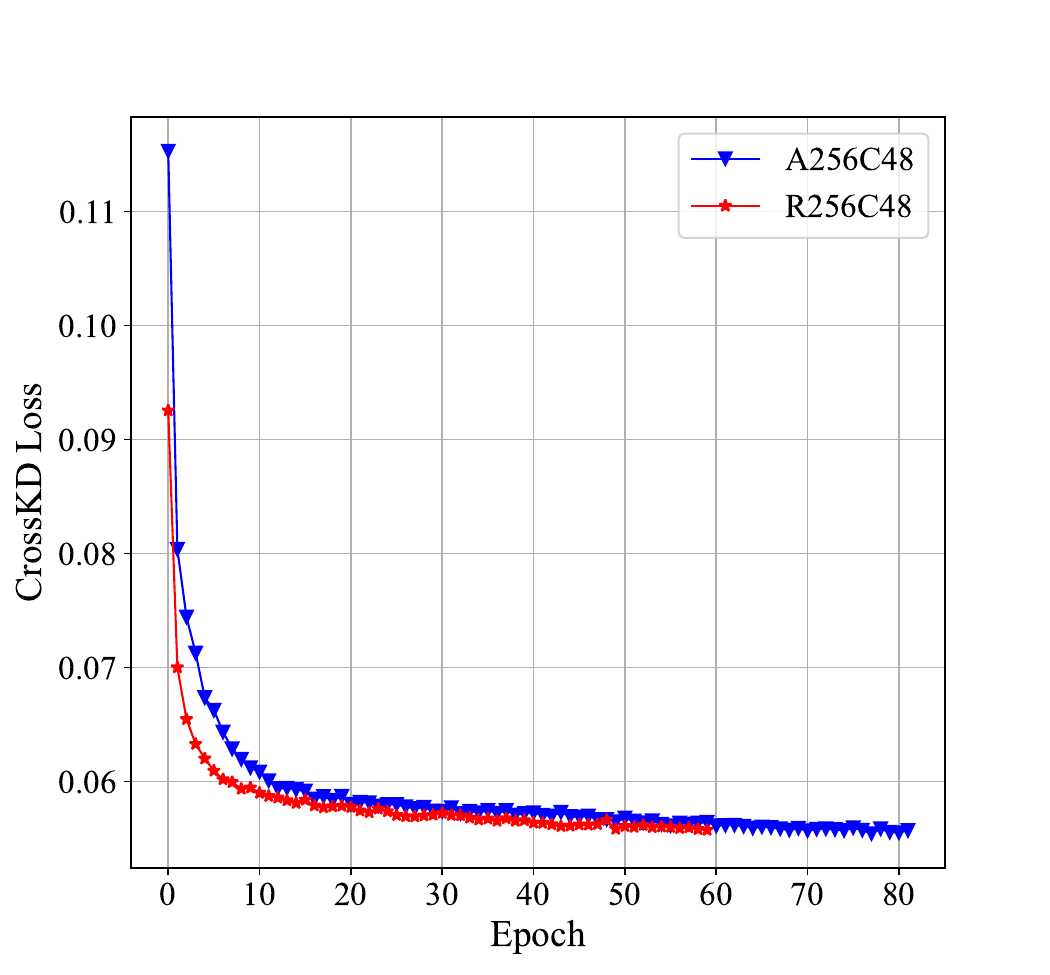}
		\captionsetup{font={footnotesize,{color=black}}}
		\caption{CrossKD Loss vs. Epoch}
		\label{fig9a}
	\end{subfigure}
	\hfill
	\begin{subfigure}[t]{0.5\columnwidth}
		\centering
		\includegraphics[width=\linewidth]{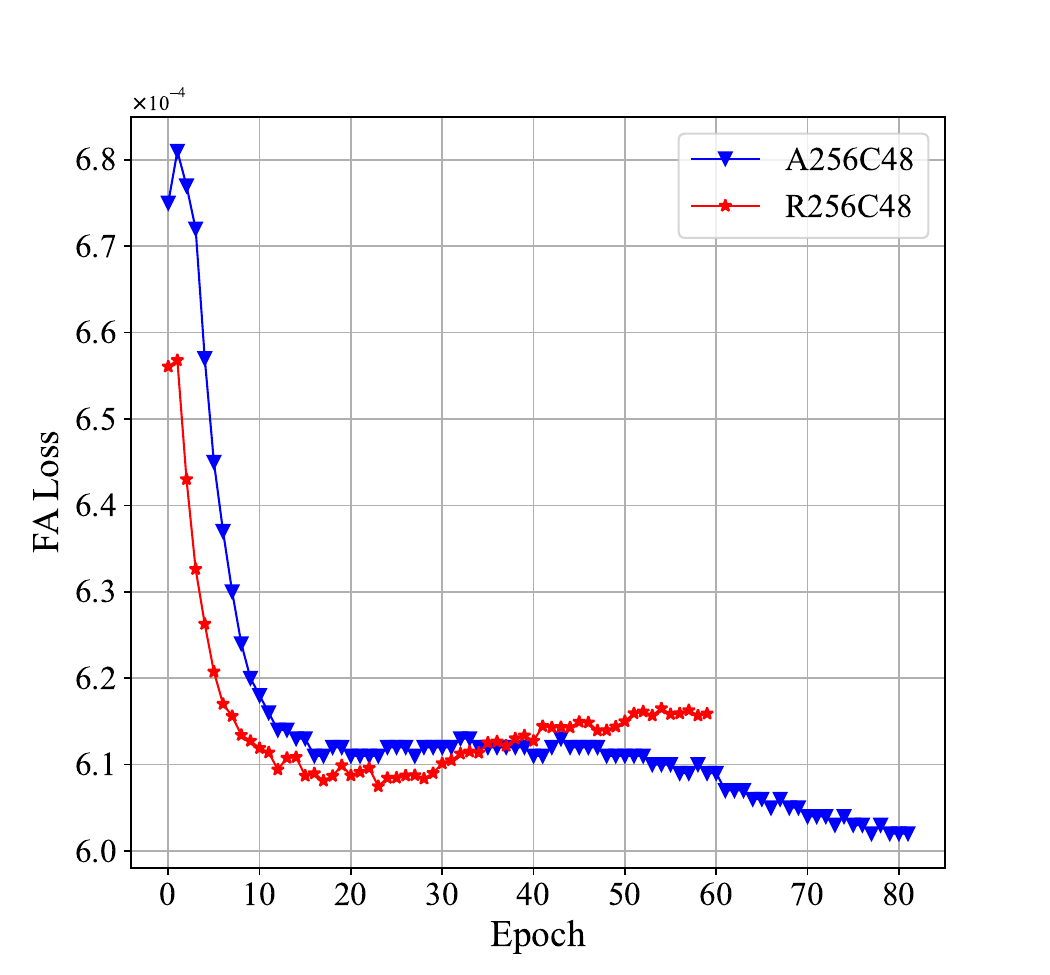}
		\captionsetup{font={footnotesize,{color=black}}}
		\caption{FA Loss vs. Epoch}
		\label{fig9b}
	\end{subfigure}
	\hfill
	\begin{subfigure}[t]{0.5\columnwidth}
		\centering
		\includegraphics[width=\linewidth]{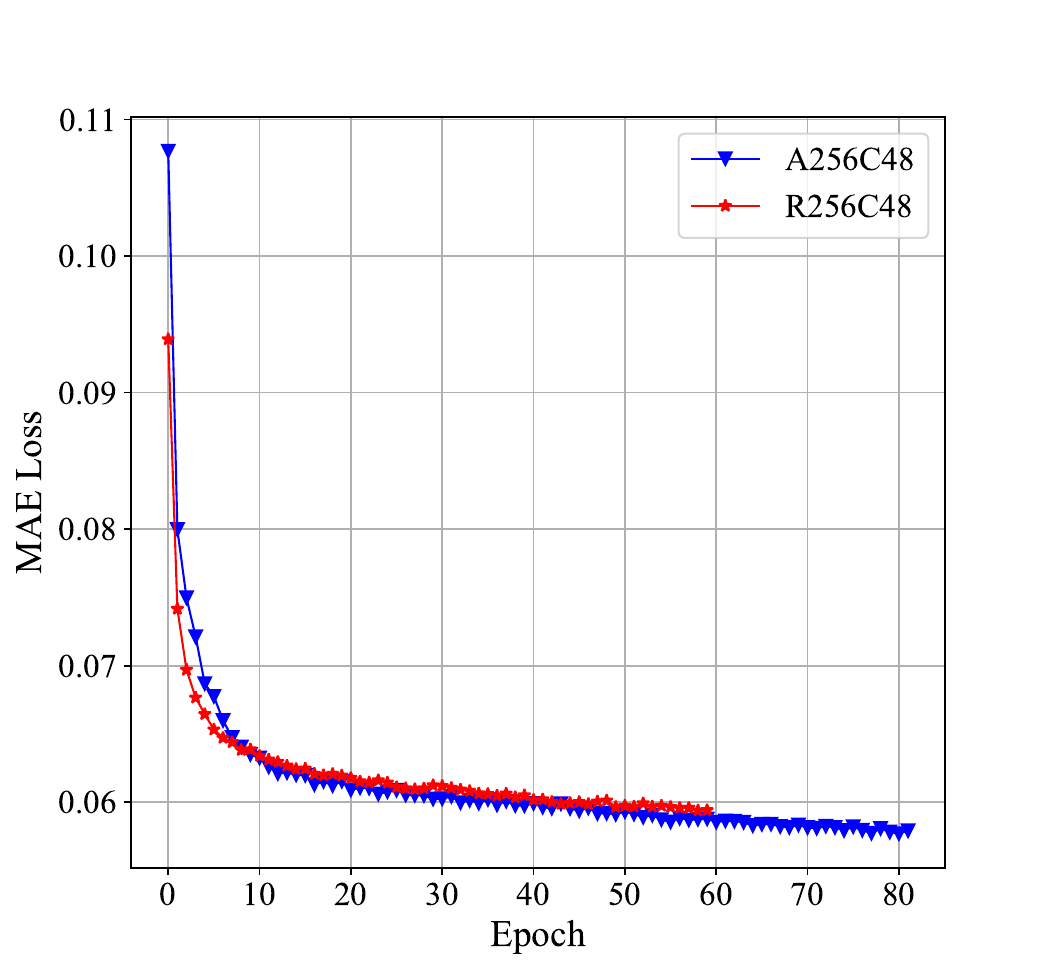}
		\captionsetup{font={footnotesize,{color=black}}}
		\caption{MAE Loss vs. Epoch}
		\label{fig9c}
	\end{subfigure}
	\hfill
	\begin{subfigure}[t]{0.5\columnwidth}
		\centering
		\includegraphics[width=\linewidth]{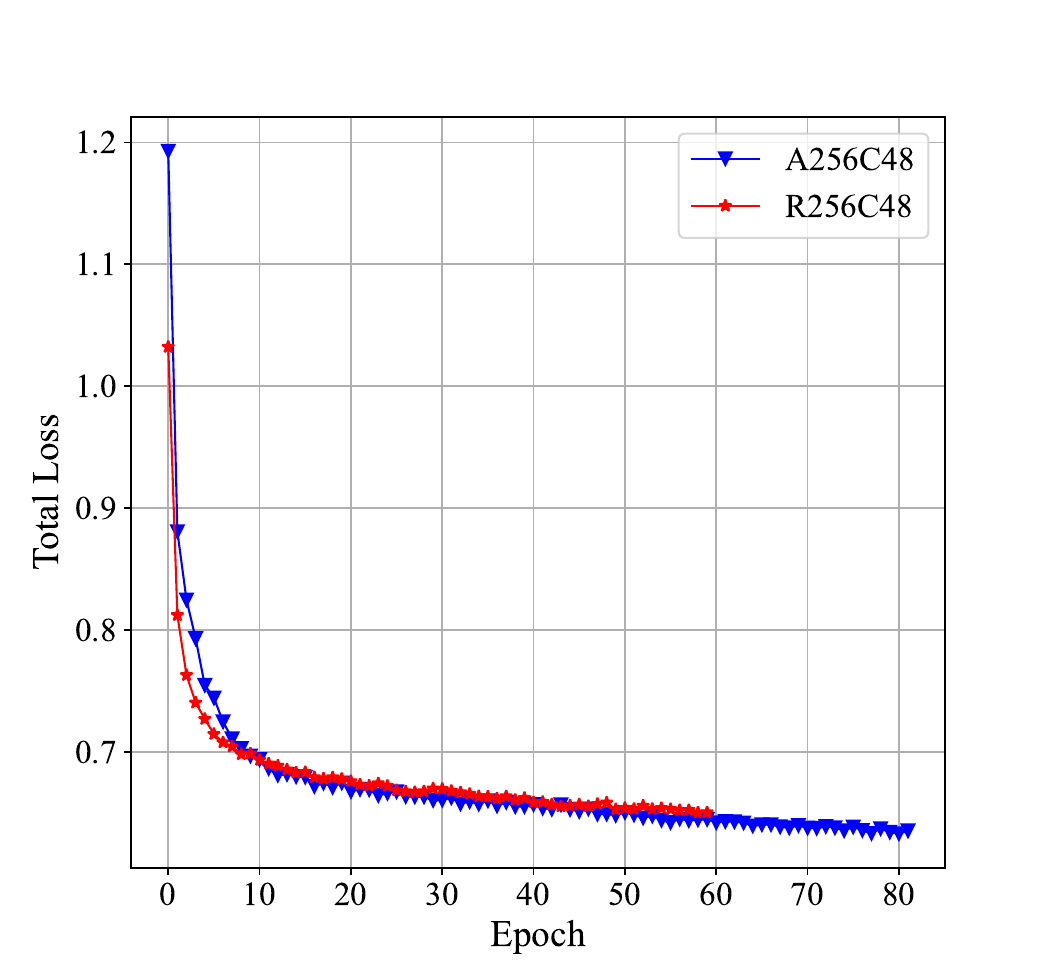}
		\captionsetup{font={footnotesize,{color=black}}}
		\caption{Total Loss vs. Epoch}
		\label{fig9d}
	\end{subfigure}
	\captionsetup{font={footnotesize,{color=black}},justification=raggedright,singlelinecheck=false}
	\caption{Visualization for the variation of statistics during training under cases of \texttt{A256C48} and \texttt{R256C48}.}
	\label{fig:loss_during_training}
	\vspace{-4mm}
\end{figure*}

We also conduct simulations on the high-resolution FFHQ-256 dataset to verify the compatibility of the proposed scheme with different resolution scenarios. As shown in Fig.~\ref{fig:performance_ffhq_M32}, we evaluate the image reconstruction PSNR and SSIM performance of different schemes under AWGN and Rayleigh channel with a compression ratio of $1/48$ across various SNR conditions. Fig.~\ref{fig6a} and Fig.~\ref{fig6b} shows the simulation in AWGN channel. Compared to the baseline \textbf{DeepJSCC-NOMA} scheme, the proposed ConvNeXt-based framework \textbf{SemNOMA} demonstrates a more pronounced advantage in high-resolution scenarios, achieving a PSNR gain of $0.3 \sim 0.4$dB, with further improvements of $0.3 \sim 0.5$dB after knowledge distillation optimization in \textbf{KD-SemNOMA}. The SSIM metric also shows consistent improvement. Compared to the orthogonal transmission scheme \textbf{SemOMA}, the proposed non-orthogonal scheme \textbf{KD-SemNOMA} achieves a performance gain of approximately $0.3 \sim 0.5$dB at the same overhead. {\color{black}Notably, the traditional \textbf{BPG+LDPC+QAM+SIC} scheme exhibits the typical cliff effect of separation-based approaches, where the reconstruction quality drops sharply when the channel quality falls below a certain threshold (around 5-6 dB SNR). In contrast, all semantic communication-based schemes demonstrate graceful performance degradation in low-SNR regimes.} Fig.~\ref{fig5a} and Fig.~\ref{fig5b} shows the simulation in Rayleigh channel.  Compared to the baseline \textbf{DeepJSCC-NOMA}, the proposed \textbf{SemNOMA} framework achieves a performance gain of $1 \sim1.5$dB, with knowledge distillation optimization providing an additional gain of approximately $0.4$dB in \textbf{KD-SemNOMA}. In terms of PSNR, the proposed \textbf{KD-SemNOMA} performs slightly worse than the orthogonal scheme \textbf{SemOMA} with the same transmission overhead in the $[0, 6]$dB range but outperforms it in the $[6, 20]$dB range. For the SSIM metric, the proposed scheme consistently outperforms \textbf{SemOMA}. {\color{black}The \textbf{BPG+LDPC+QAM+SIC} baseline shows even more severe performance degradation under Rayleigh fading conditions: the cliff effect occurs at higher SNR levels (around 10-11 dB) and decoding fails completely below these thresholds.} Overall, in high resolution and high compression ratio cases, our scheme demonstrates significant advantages over both the non-orthogonal baseline and orthogonal schemes, confirming the effectiveness of the proposed framework.
\begin{table}[htbp]
	\centering
	\setlength{\tabcolsep}{8pt}
	\caption{Effectiveness of applying CrossKD at different positions in DeepJSCC decoder at $\gamma=10$ dB}
	\label{tab:CrossKD_results}
	\begin{tabular}{c c c c c c c}
		\toprule
		$(j)$ & A32C3  & R32C3 & A256C48 & R256C48\\
		\midrule
		--    & 32.90/0.972 & 29.11/0.922 & 28.74/0.842 & 27.18/0.808  \\
		1     & 33.04/0.972 & 29.19/0.923 & 28.85/0.843 &27.24/0.809\\
		2     & 32.99/0.972 & \textbf{29.58/0.927} & 29.05/0.850 & 27.49/0.816 \\
		3     & 33.04/0.973 & 29.48/0.927 & 29.14/0.853 & 27.19/0.809  \\
		4     & \textbf{33.18/0.973} & 29.50/0.927 & 29.16/0.854 & \textbf{27.52/0.818}  \\
		5     & 32.95/0.972 & 29.14/0.921 & \textbf{29.24/0.856} & 27.41/0.816  \\
		6     & 32.92/0.972 & 29.30/0.923 & 29.01/0.846 & 27.21/0.810  \\
		7     & 32.95/0.972 & 29.47/0.926 & 29.08/0.851 & 27.38/0.813  \\
		\bottomrule
	\end{tabular}
	\vspace{-2mm}
\end{table}

As described in Section~\ref{KD}, CrossKD transfers the intermediate features of the $j$-th layer from the student model decoder to the corresponding position of the teacher model decoder for forward propagation, enhancing feature alignment in knowledge distillation (Algorithm~\ref{alg:kd}). To verify the effect of the choice of $j$ on the distillation performance, we conduct the following experiments. Cases include \texttt{A32C3} (AWGN, CIFAR-10, $\rho=1/3$), \texttt{R32C3} (Rayleigh, CIFAR-10, $\rho=1/3$), \texttt{A256C48} (AWGN, FFHQ-256, $\rho=1/48$), and \texttt{R256C48} (Rayleigh, FFHQ-256, $\rho=1/48$). As shown in Table \ref{tab:CrossKD_results}, the first row represents the student model trained without teacher model distillation, using only MAE loss. For \texttt{A32C3}, $j=4$ (upsampling layer) yields the optimal PSNR/SSIM (33.18/0.973), as upsampling features align well with AWGN’s stable conditions. For \texttt{R32C3}, $j=2$ (AF-Module layer) is the best (29.58/0.927), leveraging AF-Module’s adaptation to Rayleigh fading’s variability. In FFHQ-256 cases, $j=5$ performs best (29.24/0.856) for \texttt{A256C48} while $j=4$ performs best (27.52/0.818) for \texttt{R256C48}. These results highlight CrossKD’s ability to optimize layer-specific feature transfer, enhancing decoding performance across diverse datasets and channel conditions.

%\begin{figure*}[!t]
%	\centering
%	\begin{subfigure}[t]{1.0\columnwidth}
%		\centering
%		\includegraphics[width=\linewidth]{./Figure_wqf/figure1a.pdf}
%		\caption{PSNR vs. SNR}
%		\label{fig1a}
%	\end{subfigure}
%	\hfill
%	\begin{subfigure}[t]{1.0\columnwidth}
%		\centering
%		\includegraphics[width=\linewidth]{./Figure_wqf/figure1b.pdf}
%		\caption{SSIM vs. SNR}
%		\label{fig1b}
%	\end{subfigure}
%	\caption{Performance under AWGN channel, M=16, 2UE.}
%	\label{fig:performance_awgn_cifar10_M16}
%\end{figure*}

%\begin{figure*}[!t]
%	\centering
%	\begin{subfigure}[t]{1.0\columnwidth}
%		\centering
%		\includegraphics[width=\linewidth]{./Figure_wqf/figure2a.pdf}
%		\caption{PSNR vs. SNR}
%		\label{fig2a}
%	\end{subfigure}
%	\hfill
%	\begin{subfigure}[t]{1.0\columnwidth}
%		\centering
%		\includegraphics[width=\linewidth]{./Figure_wqf/figure2b.pdf}
%		\caption{SSIM vs. SNR}
%		\label{fig2b}
%	\end{subfigure}
%	\caption{Performance under Rayleigh channel, M=16, 2UE.}
%	\label{fig:performance_rayleigh_cifar10_M16}
%\end{figure*}

%\begin{figure*}[!t]
%	\centering
%	\begin{subfigure}[t]{1.0\columnwidth}
%		\centering
%		\includegraphics[width=\linewidth]{./Figure_wqf/figure6a.pdf}
%		\caption{PSNR vs. SNR}
%		\label{fig6a}
%	\end{subfigure}
%	\hfill
%	\begin{subfigure}[t]{1.0\columnwidth}
%		\centering
%		\includegraphics[width=\linewidth]{./Figure_wqf/figure6b.pdf}
%		\caption{SSIM vs. SNR}
%		\label{fig6b}
%	\end{subfigure}
%	\caption{Performance under AWGN channel, M=32, 2UE.}
%	\label{fig:performance_awgn_ffhq_M32}
%\end{figure*}
{\color{black}To validate the effectiveness of our knowledge distillation based training strategy, we conducted a comprehensive analysis of the loss convergence behavior during training. As shown in Fig. \ref{fig:loss_during_training}, both distillation losses ($\mathcal{L}_{\mathrm{FA}}$ and $\mathcal{L}_{\text{CrossKD}}$) exhibit smooth and stable convergence patterns across different channel conditions, indicating effective knowledge transfer from the teacher model to the student model. Importantly, the distillation losses and the reconstruction loss ($\mathcal{L}_{\text{MAE}}$) demonstrate complementary optimization behavior without conflicting trends, suggesting that the knowledge distillation process enhances rather than interferes with the primary reconstruction task. The overall loss $\mathcal{L}_S$ shows consistent and stable convergence throughout the training process, further validating the robustness of our multi-loss optimization framework and the effectiveness of the knowledge distillation approach in improving the student model's performance in non-orthogonal transmission scenarios.}
%\vspace{-3mm}
\subsection{Performance of Diffusion-Based Image Refinement for SemNOMA}
\begin{figure*}[!t]
	\centering
	\begin{subfigure}[t]{0.5\columnwidth}
		\centering
		\includegraphics[width=\linewidth]{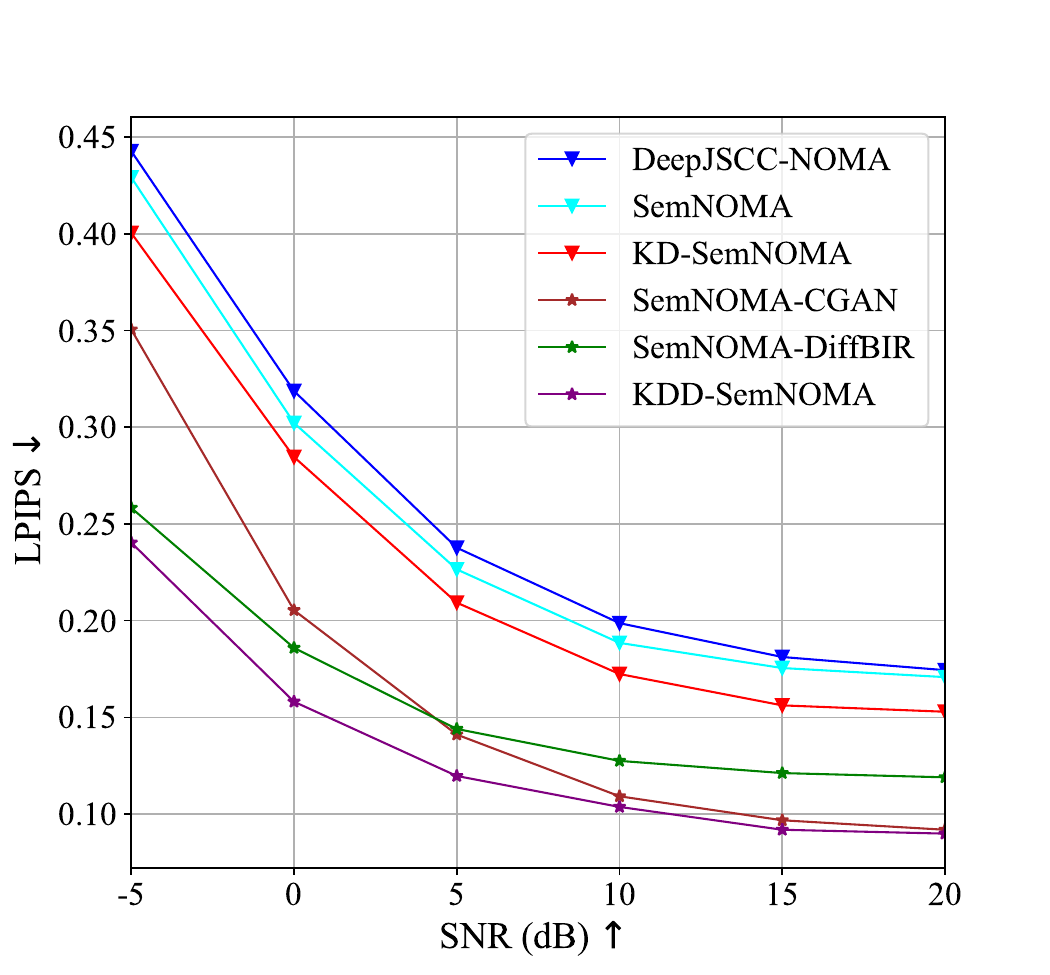}
		\captionsetup{font={footnotesize}}
		\caption{LPIPS vs. SNR (AWGN)}
		\label{fig8a}
	\end{subfigure}
	\hfill
	\begin{subfigure}[t]{0.5\columnwidth}
		\centering
		\includegraphics[width=\linewidth]{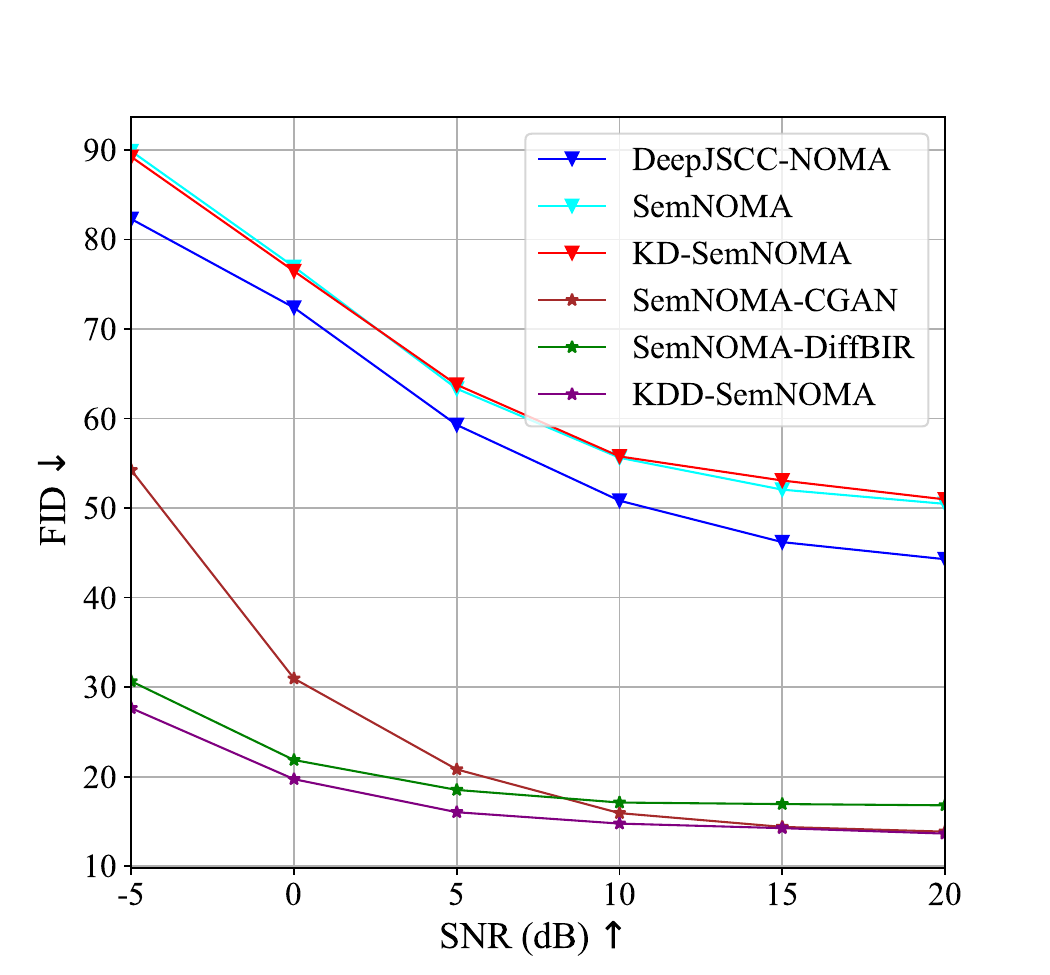}
		\captionsetup{font={footnotesize}}
		\caption{FID vs. SNR (AWGN)}
		\label{fig8b}
	\end{subfigure}
	\hfill
	\begin{subfigure}[t]{0.5\columnwidth}
		\centering
		\includegraphics[width=\linewidth]{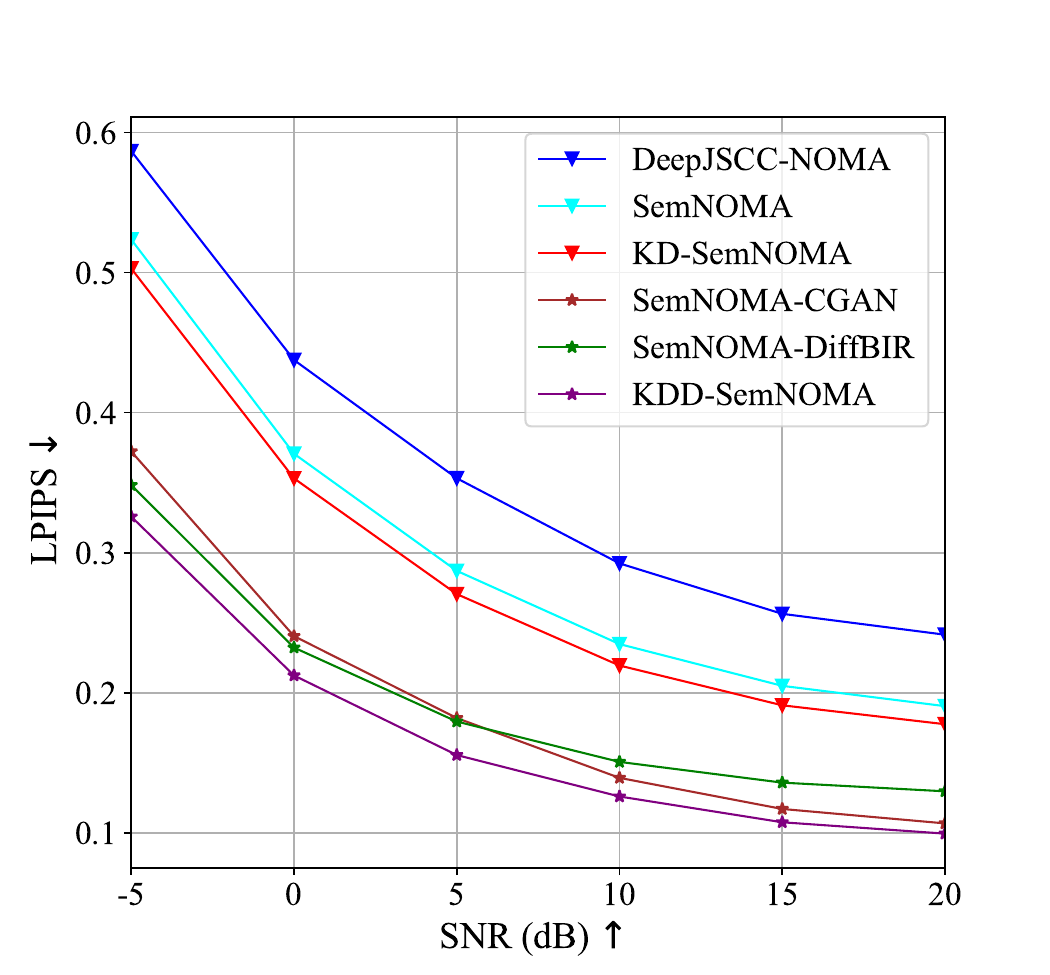}
		\captionsetup{font={footnotesize}}
		\caption{LPIPS vs. SNR (Rayleigh)}
		\label{fig7a}
	\end{subfigure}
	\hfill
	\begin{subfigure}[t]{0.5\columnwidth}
		\centering
		\includegraphics[width=\linewidth]{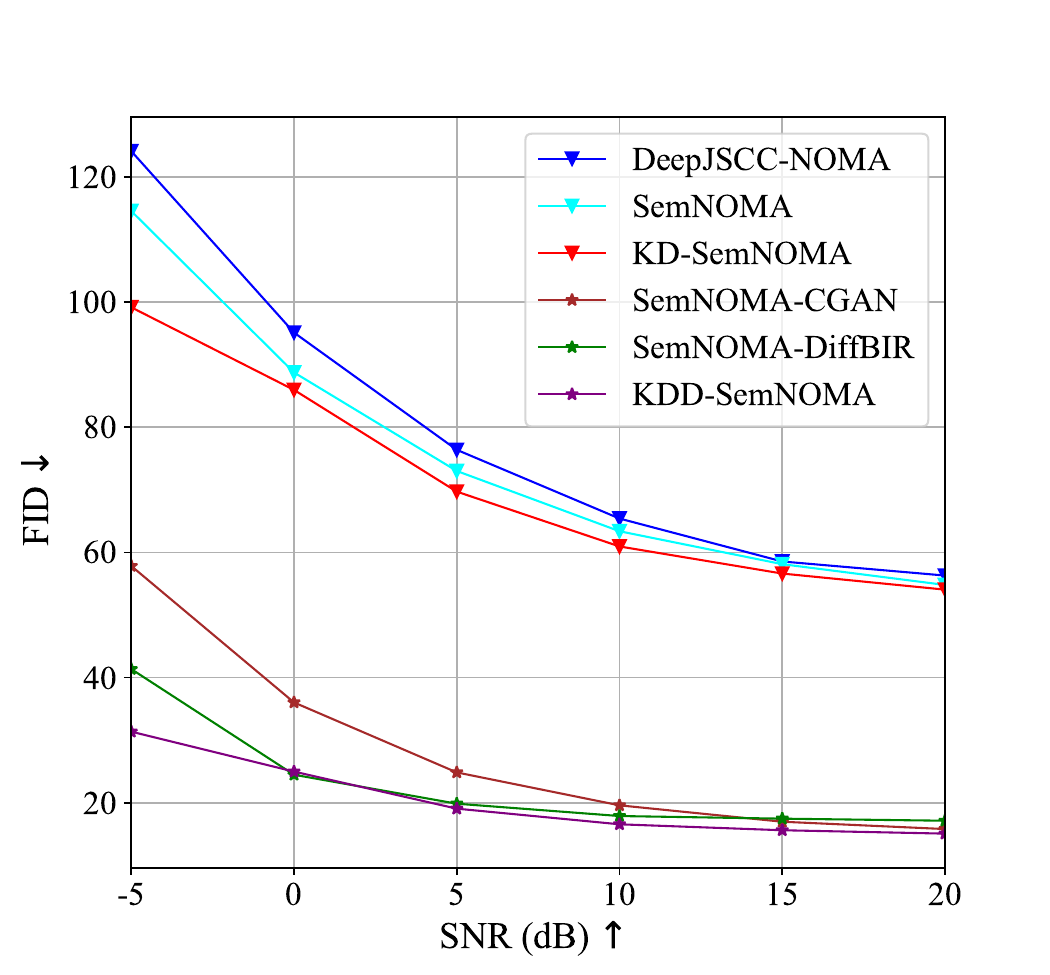}
		\captionsetup{font={footnotesize}}
		\caption{FID vs. SNR (Rayleigh)}
		\label{fig7b}
	\end{subfigure}
	\captionsetup{font={footnotesize,{color=black}}}
	\caption{Perceptual quality comparison under AWGN and Rayleigh fading channels (2UE, $\rho=1/48$). Subfigures (a) and (b) show LPIPS and FID versus SNR under AWGN channel, respectively, while subfigures (c) and (d) show LPIPS and FID under Rayleigh fading channel.}
	\label{fig:performance_lpips_fid_ffhq_M32}
	\vspace{-2mm}
\end{figure*}

In this subsection, to further improve the perceptual quality of reconstructed images, we evaluate the proposed \textbf{KDD-SemNOMA} framework. The first stage leverages knowledge distillation (Algorithm~\ref{alg:kd}) to optimize the proposed \textbf{SemNOMA} network, improving PSNR and SSIM for initial reconstructions $\mathbf{x}_{\text{init}} = \hat{\mathbf{x}}_i$ (Eq.~\ref{eq:decoding}). The second stage utilizes the diffusion model and employs DDIM sampling ($\eta = 0.5$, $T' = 100$, Algorithm~\ref{alg:diffusion}) to refine $\mathbf{x}_{\text{init}}$ into ${\mathbf{x}_0}$ (Eq.~\ref{eq:refinement_pipeline}), enhancing LPIPS and FID perceptual quality under challenging channel conditions.

Simulations are conducted under AWGN and Rayleigh fading channels with the compression ratio $\rho=1/48$, across SNR $\gamma$ ranges from -5 to 20 dB.
We compare the proposed two-stage \textbf{KDD-SemNOMA} scheme against other two-stage refinement methods {\color{black}(including Stable Diffusion based \textbf{SemNOMA-DiffBIR} and CGAN based \textbf{SemNOMA-CGAN})} and one-stage baseline schemes (\textbf{KD-SemNOMA}, \textbf{SemNOMA}, \textbf{DeepJSCC-NOMA}). As shown in Fig.~\ref{fig:performance_lpips_fid_ffhq_M32}, \textbf{KDD-SemNOMA} achieves superior LPIPS and FID across all SNR levels, with particularly notable improvements in the low SNR range of [-5, 5] dB. For instance, at $\gamma$ = 0dB under AWGN channel, \textbf{KDD-SemNOMA} significantly reduces the LPIPS and FID values compared with the one-stage method, indicating that the utilize of the diffusion model greatly improves the perceptual quality of the received image. At the same time, compared with \textbf{SemNOMA-CGAN}, \textbf{KDD-SemNOMA} reduces the LPIPS and FID values by approximately $23\%$ and $36\%$, respectively, reflecting the image generation advantages of the diffusion model over the CGAN model and its robustness to severe channel noise. {\color{black}Compared with \textbf{SemNOMA-DiffBIR}, which also utilizes the powerful generative capability of diffusion models, the two schemes adopt distinct refinement mechanisms: \textbf{SemNOMA-DiffBIR} extracts both textual and visual features from the initial estimate and uses them as conditional guidance to steer the Stable Diffusion denoising process, whereas \textbf{KDD-SemNOMA} first adds noise to the initial estimate to reduce reconstruction error and then performs reverse denoising to generate high-quality images. This enables \textbf{KDD-SemNOMA} to achieve superior LPIPS and FID performance across different SNR regimes.}

The proposed scheme \textbf{KDD-SemNOMA} outperforms \textbf{SemNOMA-CGAN}, as diffusion models capture richer image distributions and recover finer details, leading to lower LPIPS and FID scores. Unlike GAN-based methods, which often struggle with mode collapse and artifacts in low-SNR scenarios, the iterative denoising of the diffusion model (Eq.~\ref{eq:reverse_sampling}) effectively mitigates channel-induced distortions, especially at $\gamma \in [-5, 5]$ dB. Additionally, the accelerated sampling with \(T' = 100\) (compared to DDPM's \(T \approx 1000\)) ensures computational efficiency \cite{songdenoising}. These results underscore the efficacy of combining knowledge distillation and diffusion-based refinement for high-fidelity multi-user image reconstruction.

\begin{figure*}[!t]
	\centering
	\includegraphics[width=1\linewidth]{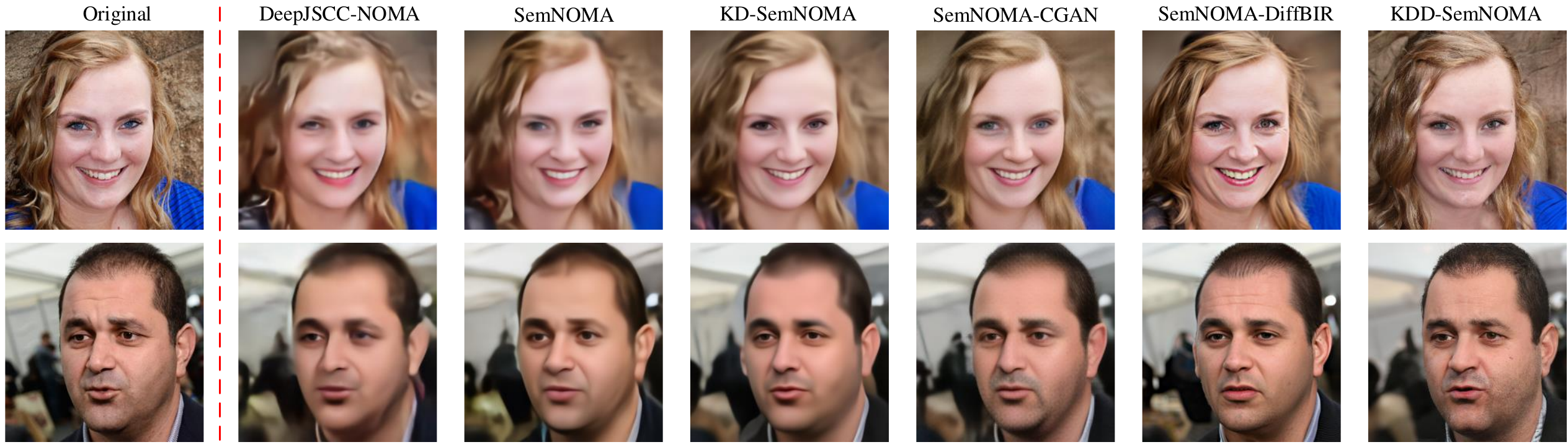}
	\captionsetup{font={footnotesize,{color=black}},justification=raggedright,singlelinecheck=false}
	\caption{Examples of reconstructed images under AWGN channel with $\gamma=0$ dB.}
	\label{fig:example of awgn reconstruction images snr=0db}
	\vspace{-2mm}
\end{figure*}
\begin{figure*}[!t]
	\centering
	\includegraphics[width=1\linewidth]{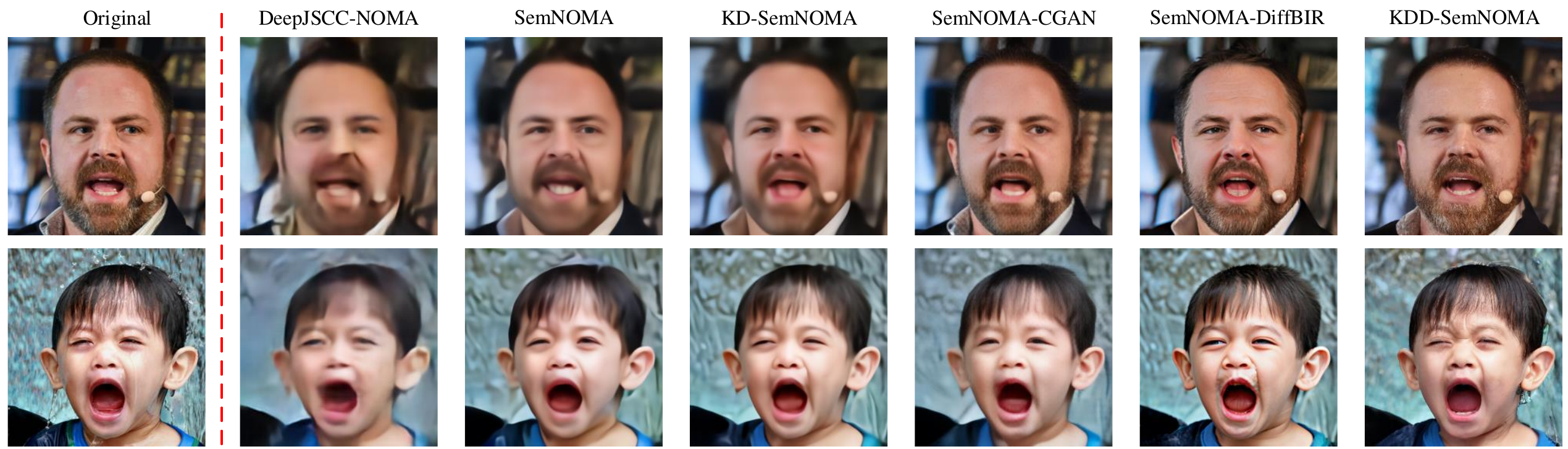}
	\captionsetup{font={footnotesize,{color=black}},justification=raggedright,singlelinecheck=false}
	\caption{Examples of reconstructed images under Rayleigh channel with $\gamma=0$ dB.}
	\label{fig:example of rayleigh reconstruction images snr=0db}
	\vspace{-2mm}
\end{figure*}

We provide examples of reconstructed images from FFHQ-256 datasets in Fig.~\ref{fig:example of awgn reconstruction images snr=0db} and Fig.~\ref{fig:example of rayleigh reconstruction images snr=0db}, showcasing visual quality at $\gamma=0$ dB under AWGN and Rayleigh fading channels, respectively. While the one-stage method \textbf{KD-SemNOMA} achieve competitive PSNR and SSIM for initial reconstructions $\mathbf{x}_{\text{init}} = \hat{\mathbf{x}}_i$, its images suffer from noise-induced degradation at $\gamma=0$ dB, losing semantic details such as facial contours. {\color{black}The two-stage refinement schemes generally improve perceptual quality}. However, the GAN-based generation method \textbf{SemNOMA-CGAN} often introduces artifacts and blurred textures due to mode collapse in low-SNR conditions, particularly under Rayleigh fading. {\color{black}In comparison, both diffusion-based two-stage methods, \textbf{SemNOMA-DiffBIR} and \textbf{KDD-SemNOMA}, recover finer semantic details and exhibit enhanced visual coherence. Nevertheless, \textbf{KDD-SemNOMA} more faithfully preserves original texture details and facial features, benefiting from its targeted error contraction in the forward diffusion process (Eq.~\ref{eq:error_contraction}) and iterative denoising guided by diffusion priors (Eq.~\ref{eq:reverse_sampling}).} %Visualizations demonstrate that \textbf{KDD-SemNOMA} restores finer facial details, such as hair textures and eye clarity, outperforming \textbf{SemNOMA-CGAN} by producing more natural and perceptually coherent images \cite{yue2024difface}.

%\begin{figure}[!t]
%	\centering
%	\includegraphics[width=3.5in]{./Figure_wqf/figure8a.pdf}
%	\caption{LPIPS, AWGN Channel, 2UE, M=32, FFHQ256}
%	\label{fig8a}
%\end{figure}

%\begin{figure}[!t]
%	\centering
%	\includegraphics[width=3.5in]{./Figure_wqf/figure8b.pdf}
%	\caption{FID, AWGN Channel, 2UE, M=32, FFHQ256}
%	\label{fig8b}
%\end{figure}

%\begin{figure}[!t]
%	\centering
%	\includegraphics[width=3.5in]{./Figure_wqf/figure7a.pdf}
%	\caption{LPIPS, Rayleigh Channel, 2UE, M=32, FFHQ256}
%	\label{fig7a}
%\end{figure}

%\begin{figure}[!t]
%	\centering
%	\includegraphics[width=3.5in]{./Figure_wqf/figure7b.pdf}
%	\caption{FID, Rayleigh Channel, 2UE, M=32, FFHQ256}
%	\label{fig7b}
%\end{figure}
%\vspace{-5.1mm}
\subsection{Ablation Study}
To validate the effectiveness of the knowledge distillation optimization scheme in enhancing model performance, we conduct ablation studies using following datasets (CIFAR-10, FFHQ-256), compression ratios (1/3, 1/48), and channel conditions (AWGN, Rayleigh) at $\gamma=10$ dB. Results are presented in Table~\ref{tab:distillation components}, with cases defined as follows: \texttt{A32C3} (AWGN, CIFAR-10, $\rho=1/3$), \texttt{R32C3} (Rayleigh, CIFAR-10, $\rho=1/3$), \texttt{A256C48} (AWGN, FFHQ-256, $\rho=1/48$), and \texttt{R256C48} (Rayleigh, FFHQ-256, $\rho=1/48$). The first row represents the baseline where only reconstruction loss is used, indicating that the student model is trained without distillation from the teacher model. The subsequent rows demonstrate the performance of the student model trained with different distillation algorithms. Specifically, the second and third rows confirm that both FA and CrossKD distillation improve model performance. The final row indicates that the best performance is achieved when FA and CrossKD are applied simultaneously.
Notably, performance gains are consistent across datasets and channels. These findings align with results in Fig.~\ref{fig:performance_awgn_cifar10} and Fig.~\ref{fig:performance_rayleigh_cifar10} for CIFAR-10 and Fig.~\ref{fig:performance_ffhq_M32} for FFHQ-256.
\begin{table}[htbp]
	\centering
	\setlength{\tabcolsep}{5pt}
	%\captionsetup{font={footnotesize}}
	\caption{Effects of different distillation components on the performance of the student model at $\gamma=10$ dB}
	\label{tab:distillation components}
	\begin{tabular}{c c c c c c c}
		\toprule
		FA & CrossKD & A32C3  & R32C3 & A256C48 & R256C48 \\
		\midrule
		-- & -- & 32.90/0.972 & 29.11/0.921 & 28.74/0.842 & 27.18/0.808   \\
		-- & $\checkmark$ & 33.13/0.973 & 29.23/0.922 & 28.92/0.848  & 27.30/0.811    \\
		$\checkmark$ & -- & 33.14/0.973 & 29.36/0.932 & 29.03/0.850 & 27.41/0.814   \\
		$\checkmark$ & $\checkmark$ & \textbf{33.18/0.973} & \textbf{29.58/0.927} & \textbf{29.24/0.856} & \textbf{27.52/0.818}   \\
		\bottomrule
	\end{tabular}
	\vspace{-2mm}
\end{table}
%\vspace{-2.8mm}
\section{Conclusion} \label{Conclusion}
This paper proposed the KDD-SemNOMA scheme, a pioneering framework for wireless image semantic NOMA transmission. By integrating adaptive channel-aware encoding, knowledge distillation optimization, and diffusion model-based generative refinement, our approach effectively addressed the critical challenges of semantic feature interference and perceptual quality degradation under bandwidth-constrained multi-user scenarios. Our proposed KDD-SemNOMA leveraged a ConvNeXt-based architecture with enhanced AF-Module to ensure robust performance over AWGN and Rayleigh fading channels. At the first stage, knowledge distillation strategy leveraged interference-free orthogonal transmission teacher model to optimize SemNOMA student model training, yielding PSNR and SSIM gain without additional inference complexity. At the second stage, the diffusion model-based refinement harnessed generative priors to elevate perceptual quality, transforming initial reconstructions into high-fidelity images. Experimental results demonstrate KDD-SemNOMA’s superiority over state-of-the-art baselines, simultaneously advancing pixel-level fidelity and perceptual quality. Future work will extend the scenario from uplink to downlink NOMA, explore scalability to larger user populations, adjust the compression rate based on real-time channel state information, and integration of multimodal data.

%\vfill
%\vspace{-3.5mm}
\bibliographystyle{IEEEtran}
\bibliography{IEEEabrv,wqf_bib}

\end{document}